\documentclass[12pt]{article}
 \usepackage{graphicx,psfrag,epsfig}
\newcommand{\beq}{\begin{equation}}
\newcommand{\eeq}{\end{equation}}

\newcommand{\bea}{\begin{eqnarray}}
\newcommand{\eea}{\end{eqnarray}}

\newcommand{\ea}{{\it et al.}}

\def\lsim{\mathrel{\vcenter{\hbox{$<$}\nointerlineskip\hbox{$\sim$}}}}
\def\gsim{\mathrel{\vcenter{\hbox{$>$}\nointerlineskip\hbox{$\sim$}}}}
\begin{document}
\begin{flushright}
UMD-PP-09-055\\
October, 2009\
\end{flushright}
\vspace{0.3in}
\begin{center}
{\LARGE \bf  TeV Scale Inverse Seesaw in SO(10) and
Leptonic Non-Unitarity Effects}\\
\vspace{0.5in}
{\bf P. S. Bhupal Dev and R. N. Mohapatra\\}
\vspace{0.2in} {\sl Maryland Center for Fundamental Physics and
Department of Physics,\\ University of Maryland, College Park, MD 20742, USA}
\end{center}
\vspace{0.5in}
\begin{abstract}
 We show that a TeV scale inverse seesaw model for neutrino masses
can be realized within the framework of a supersymmetric $SO(10)$
model consistent with gauge coupling unification and observed
neutrino masses and  mixing. We present our expectations for
non-unitarity effects in the leptonic mixing matrix some of which 
are observable at future neutrino
 factories as well as the next generation searches for lepton flavor
violating processes such as $\mu\to e+\gamma$. The model has TeV scale
$W_R$ and $Z'$ bosons
which are accessible at the Large Hadron Collider.
\end{abstract}
\section{Introduction}
 A precise understanding of the origin of observed neutrino masses and mixing
is one of the major goals of particle physics right now. A simple
 paradigm for understanding the smallness of the  masses is the seesaw
mechanism~\cite{type1} where one introduces three Standard Model (SM)
singlet right-handed (RH) neutrinos with Majorana masses $M_N$, which mix with
left-handed (LH) ones via the Yukawa coupling $\overline{L}HN$. The resulting
formula for light  neutrino masses is given by
${\cal M}_\nu=-M_DM^{-1}_NM^T_D$, where $M_D$ is the Dirac neutrino mass.
Since the SM does not restrict the
Majorana mass $M_N$, we could choose  this to be much larger than the weak
scale thereby providing a natural way to understand the tiny neutrino masses.
This is called the type I seesaw. There are several variations of this
mechanism  where one replaces the RH neutrino by either a SM triplet Higgs
field (type II seesaw)~\cite{type2} or SM triplet of fermions (called type III
 seesaw)~\cite{type3}. A great deal of attention has been devoted to testing
these ideas. As far as the type I seesaw is concerned, the prospects of
 testing this depends on the scale $M_N$ as well as any associated physics
that comes with it at that scale. It can be accessible to current and
 future collider experiments if the scale is not far above a  TeV. A different
way to test the type I seesaw mechanism follows from the observation
 that this mechanism involves the mixing of the LH neutrinos with SM singlet
heavy neutrinos as a result of which there would in general be
 violation of unitarity of the Pontecorvo-Maki-Nakagawa-Sakata (PMNS) mixing
matrix that describes only the
mixing of the three light neutrinos. One could contemplate searching for
 such effects in oscillation experiments~\cite{antjhep}. However, in the
type I seesaw case, the resulting mixing effects are of order
 $\frac{m_\nu}{M_N}$ and since neutrino masses are in the sub-eV range,
such non-unitarity effects are too small to be observable for generic
 high scale seesaw models. This would also be true with TeV mass RH
neutrinos unless there are cancellations to get small neutrino masses from large Dirac masses using symmetries (see for example cases with~\cite{smirnov}). 
We note that the non-unitarity effects
are also there in the type III seesaw case but not in the type II case even 
though they have other interesting effects such as LFV processes~\cite{bigjhep}. 

  Since the testability of seesaw is intimately related to the
magnitude of the seesaw scale, a key  question of interest is whether there
  could be any theoretical guidelines for the seesaw scale. In such a case,
the searches for seesaw effects in experiments could then be used to test
  the nature of physics beyond the standard model. It is well-known that~
\cite{review} the simplest grand unified theory (GUT) realizations of the
  seesaw mechanism are based on the $SO(10)$ group which automatically
predicts the existence of the RH neutrinos (along with the SM fermions)
required by the seesaw mechanism.   An advantage of GUT embedding of the
seesaw mechanism is that the constraints of GUT symmetry tends to relate the
Dirac neutrino mass $M_D$ to the charged fermion masses thereby making a
prediction for the seesaw scale $M_N$ from observations. For type I seesaw
GUT embedding, typical values for the $M_N$ are very
  high (in the range of $10^{10}$ - $10^{14}$ GeV). This makes both the
collider as well as non-unitarity probes of seesaw impossible. The key feature
  that leads to such restrictions in type I seesaw case is the close link
between the $B-L$ breaking RH neutrino mass and the smallness of the LH
neutrino masses.

  A completely different realization of the seesaw mechanism is the so-called
inverse seesaw mechanism~\cite{inverse}, where instead of one set of
  three SM singlet fermions, one introduces two sets of them $N_i,~ S_i$
($i=1,2,3$). In the context of $SO(10)$ models, since one of the two
sets can be identified with the SM singlet neutrino in the $SO(10)$
{\bf 16}-representation containing matter, the others would have
to be a separate set of three $SO(10)$ singlet fermions. Due to the
existence of the second set of singlet fermions (and perhaps
additional gauge symmetries e.g. $SO(10)$), the neutrino mass
formula in these models has the form
  \begin{eqnarray}
m_\nu \simeq M_DM_N^{-1}\mu\left(M_N^T\right)^{-1}M_D^T\equiv F\mu F^T
\label{eq:neutrino}
\end{eqnarray}
  where $\mu$ breaks the lepton number. Because of the presence of this new
mass scale in this theory, the seesaw scale $M_N$  can be very
close to a TeV
 even for ``large'' Dirac masses. This makes the tests of this possibility in
colliders much more feasible. In fact it  has recently been argued
  that~\cite{malinsky} the inverse seesaw scenario can also lead to
non-negligible non-unitarity effects which can be accessible at the
  future long-baseline neutrino oscillation experiments. There are
 also significant lepton flavor violation (LFV) effects in these models as noted
many years ago in Ref.~\cite{pila}. These possibilities have generated
a great deal of interest in the inverse seesaw models in recent
days~\cite{many}. Our effort in this paper focuses on possible
grand unification of inverse seesaw models.
Similar unification studies have been performed in Ref.~\cite{fukujhep}, but
they have not addressed the non-unitarity issues.

 An interesting question is whether such models are necessarily compatible
with grand unification when the seesaw scale is in the TeV range
and if so what kind of non-unitarity effects they predict. We find
that it is indeed possible to embed the TeV scale inverse seesaw
models within a simple
 $SO(10)$ framework consistent with gauge coupling unification and realistic
fermion masses. The $SO(10)$ symmetry helps to reduce the number
of parameters in the inverse seesaw matrix, once we require
degeneracy of the TeV scale RH neutrinos to have successful
resonant leptogenesis. Within this set of assumptions, we present
our expectations for the non-unitarity effects as well as
consequences for lepton flavor violation which are in the testable
range in future experiments.

This paper is organized as follows: In Section 2, we describe the general
framework of the inverse seesaw model and
its embedding into a generic supersymmetric $SO(10)$ theory. In Section 3,
we analyze the non-unitarity predictions of the inverse seesaw model. In
Section 4, we investigate a  specific $SO(10)$ breaking chain and obtain the
gauge coupling unification with TeV scale left-right symmetry with unification
scale consistent with proton decay bounds.
In Section 5, we analyze the renormalization group (RG) evolution of the
Yukawa couplings and obtain the running
masses for quarks and leptons at the unification scale to check that our model
leads to realistic fermion masses. In
Section 6, we determine the Dirac neutrino mass matrix using the results of
Section 5. In Section 7, we study the implication of our model on non-unitarity
effects and its phenomenological consequences. A brief summary of the results
is presented in Section 8. In Appendix A, we have given the expressions for
the masses of the $SO(10)$ Higgs multiplets in our model, and
in Appendix B, we have derived the RG equations for the quark  and lepton
masses and the CKM mixing elements in the context of a supersymmetric
left-right model.
\section{The Inverse Seesaw Model}
The inverse seesaw scheme was originally suggested~\cite{inverse} for theories
which lack the representation
required to implement the canonical seesaw, such as the superstring models.
As noted in the introduction, the implementation of the
inverse seesaw requires the addition of three extra SM gauge singlets
$S_i$ coupled to the RH neutrinos $N_i$ through the lepton number conserving
couplings of the type $\overline{N}S$ while the traditional
RH neutrino Majorana mass term is forbidden by extra symmetries. The lepton
number is broken only by the self
coupling term $\mu S^2$. The mass part of the neutrino sector
Lagrangian in the flavor basis is given by
\begin{eqnarray}
{\cal L}_{\rm mass}=(\overline{\nu}M_D N+\overline{N}M_N S+{\rm h.c.})
+S\mu S
\end{eqnarray}
where $\mu$ is a complex symmetric $3\times 3$ matrix (with dimension of
mass), and $M_D$ and $M_N$ are generic $3\times 3$ complex matrices
representing the
Dirac mass terms in the $\nu$ - $N$ and $N$ - $S$ sectors
respectively. In the basis $\{\nu,N,S\}$, the $9\times 9$ neutrino mass matrix
becomes
\begin{eqnarray}
{\cal M}_\nu = \left(\begin{array}{ccc}
0 & M_D & 0\\
M_D^T & 0 & M_N\\
0 & M_N^T & \mu
\end{array}\right)
\label{eq:big}
\end{eqnarray}
The LH neutrinos can be made very light (sub-eV scale), as required by the
oscillation data, even for a low $M_N$,
much smaller than the unification scale ($M_N\ll M_G$), provided $\mu$ is
sufficiently small, $\mu\ll M_N$, as the lepton number breaking scale $\mu$ is
decoupled from the RH neutrino mass scale.
 Assuming $\mu\ll M_\nu^D\ll M_N$ (with $M_N\sim$ TeV), the structure of the
light neutrino Majorana mass term at the leading order in $M_DM_N^{-1}$ is
given by Eq.~(\ref{eq:neutrino}),
where $F=M_DM_N^{-1}$ is a complex $3\times 3$ matrix. We note that in the
limit $\mu\to 0$ which corresponds to
the unbroken lepton number, we have massless LH neutrinos as in the SM.
In reality, a small non-vanishing $\mu$ can be viewed as a slight breaking of
a global $U(1)$ symmetry; hence, the smallness of $\mu$ is natural, in the
't Hooft sense~\cite{thooft}, even though there is no dynamical understanding
of this smallness.

The generic form of the inverse seesaw matrix in Eq.~(\ref{eq:big}) has more
parameters than the usual type I seesaw. However, if we embed this theory into
a grand unified theory such as $SO(10)$, that will help in reducing the
parameters as we show below.
 In order to embed the inverse see-saw mechanism into a generic $SO(10)$
theory, we have to break the $B-L$ symmetry
by using a $\mathbf {16}~\oplus~\overline{\mathbf {16}}$ pair rather than the
$\mathbf {126}~\oplus~\overline{\mathbf {126}}$ pair of Higgs representation.
All the SM fermions are accommodated in a single
${\bf 16}_F$ representation of $SO(10)$ and we use three copies of
${\bf 16}_F^i$ for three generations. For each of
them, we add a gauge singlet fermion ${\bf 1}_F^i$ to play the role of $S_i$.
We assume more than
one copy of ${\bf 10}_H$ Higgs multiplets in order to have a realistic
fermionic spectrum.

The $SO(10)$ invariant renormalizable Yukawa superpotential is given by
\begin{equation}
W_Y = h^a_{ij}\mathbf {16}_F^i\mathbf {16}_F^j\mathbf {10}_H^a + f_{ijk}\mathbf {16}_F^i \mathbf
{1}_F^j\overline{\mathbf {16}}_H^k + \mu_{ij}\mathbf 1_F^i\mathbf 1_F^j
\end{equation}
After the $B-L$ symmetry breaking, we get the neutrino mass matrix in
Eq.~(\ref{eq:big})  with $M_D=h v_u$ and $M_N~=~f \bar{v}_R$,
 where $v_u$ is the vacuum expectation value (VEV) of one (or, a linear
combination) of the ${\bf 10}_H$'s and $\bar v _R$ the VEV of the
$\overline {\bf 16}_H$.
In a  typical TeV-scale scenario with $v_u\sim 100$ GeV (electroweak scale)
and $\bar{v}_R\sim$ TeV, assuming $\mu\ll v_u < \bar{v}_R$, we find the
lightest neutrino mass from Eq.~(\ref{eq:neutrino}) in a one generation theory to be
\begin{equation}
m_\nu \simeq \mu\left(\frac{hv_u}{f\bar{v}_R}\right)^2
\end{equation}
and the two other heavy eigenstates with mass of order $f\bar{v}_R$. Thus, we
can get sub-eV light neutrino mass for $\mu\sim$ keV. Since this is a
supersymmetric theory, such small values do not receive radiative corrections
and keep the model natural. In the following section, we consider three
generations which then results in the non-unitarity effect.

It is important to note that in our model, we do not need to impose a discrete 
$R$-parity to our matter fermions, unlike the usual ${\bf 16}_H$ $SO(10)$ models discussed 
in literature, in order to prevent fast proton decay via dimension-4 operators 
of the type $\frac{1}{M}{\bf 16}_F{\bf 16}_F{\bf 16}_F{\overline{\bf 16}}_H$ 
because 
these operators are already suppressed by a factor $\frac{\langle 
\overline{\bf 16}_H
\rangle}{M_{\rm Pl}}\sim 10^{-15}$ for a low-scale $B-L$ breaking with 
$\langle \overline{\bf 16}_H\rangle 
\sim$ TeV. 
\section{Non-unitarity Effects}
The $3\times 3$ light neutrino mass matrix in Eq.~(\ref{eq:neutrino}) can
be diagonalized by a unitary transformation:
\begin{eqnarray}
U^\dagger m_\nu U^* = \widehat{m}_\nu = {\rm diag}(m_1,m_2,m_3)
\label{eq:oldpmns}
\end{eqnarray}
where $U$ is the standard PMNS matrix.
Since the above diagonalization of $m_\nu$ does not diagonalize the
matrices $M_N$ and $\mu$, there will be off-diagonal mixing between the
different light neutrinos even after diagonalization of $m_\nu$ due to their
mixing with the heavy neutrinos. In other words,
in the basis where the charged-lepton mass matrix is diagonal, $U$ is only a
part of the full mixing matrix responsible for neutrino
oscillations.  We have to examine the full $9\times 9$ unitary matrix $V$
which diagonalizes the mass matrix ${\cal M}_\nu$ given by Eq.~(\ref{eq:big}):
\begin{eqnarray}
V^\dagger {\cal M}_\nu V^* = \widehat{{\cal M}}_\nu =
{\rm diag}(m_i,m_{N_j},m_{\tilde{N}_k})~~~~~~(i,~j,~k=1,2,3)
\end{eqnarray}
We can decompose $V$ into the blocks
\begin{eqnarray}
V=\left(\begin{array}{cc}
V_{3\times 3} & V_{3\times 6}\\
V_{6\times 3} & V_{6\times 6}
\end{array}\right)
\end{eqnarray}
  Then the upper-left sub-block $V_{3\times 3}$ will represent the full
(non-unitary) PMNS mixing matrix. For a TeV-scale $M_N$ and a reasonably
small  $\mu$, it is sufficient to consider only up to the leading order in
$F$. Then the new PMNS matrix becomes~\cite{kanaya}
\begin{eqnarray}
{\cal N}\equiv V_{3\times 3}\simeq \left(1-\frac{1}{2}FF^\dagger\right)U
\label{eq:eta}
\end{eqnarray}
 In the commonly used parametrization~\cite{alt}, ${\cal N}=(1-\eta)U$, and
hence, all the non-unitarity effects are determined by the Hermitian matrix
  $\eta\simeq \frac{1}{2}FF^\dagger$ which depends only on the mass ratio
$F=M_DM_N^{-1}$ and not on the parametrization of the PMNS matrix.

 The LH neutrinos entering the charged-current interactions of the SM now
become superpositions of the nine mass eigenstates
$(\hat{\nu}_i,N_i,\tilde{N}_i)$ and at the leading order in $F$,
\begin{eqnarray}
\nu \simeq {\cal N}\hat{\nu}+{\cal K}P
\end{eqnarray}
 where ${\cal K}\equiv V_{3\times 6}\simeq (0,F)V_{6\times 6}$ and $P=(N_1,N_2,N_3,\tilde{N}_1,\tilde{N}_2,\tilde{N}_3)$.
Then the charged-current Lagrangian in the mass basis is given by
\begin{eqnarray}
 {\cal L}_{\rm CC}&=&-\frac{g}{\sqrt 2}\overline{l}_L\gamma^\mu \nu W_\mu^-
+{\rm h.c.}\simeq -\frac{g}{\sqrt 2}\overline{l}_L\gamma^\mu
({\cal N}\hat{\nu}+{\cal K}P)W_\mu^-+{\rm h.c.}
\end{eqnarray}
This mixing between the doublet and singlet components in the charged-current
sector has several important phenomenological consequences, as listed below:
\begin{enumerate}
\item The flavor and mass eigenstates of the LH neutrinos are now connected by
a non-unitary mixing matrix ${\cal N}=(1-\eta)U$, where the non-unitarity
effects entering different neutrino oscillation channels are measured by the
parameter $\eta$. In particular, the $CP$-violating effects in the leptonic
sector will now be governed by the PMNS matrix ${\cal N}$ instead of $U$
through the Jarlskog invariant~\cite{jarlskog}
\begin{eqnarray}
J_{\alpha\beta}^{ij} = {\rm Im}\left({\cal N}_{\alpha i}{\cal N}_{\beta j}
{\cal N}^*_{\alpha j}{\cal N}^*_{\beta i}\right)
\label{eq:jarl1}
\end{eqnarray}
where the indices $\alpha\neq \beta$ run over $e,~\mu$ and $\tau$, while
$i\neq j$ can be 1, 2 and 3. In the standard PMNS parametrization of $U$
by the three mixing angles $\theta_{ij}$ and the Dirac $CP$-phase $\delta$,
one can expand Eq.~(\ref{eq:jarl1}) up to second order in $\eta_{\alpha\beta}$
and $s_{13}\equiv \sin\theta_{13}$ (assuming those to be small) to obtain
\begin{eqnarray}
J_{\alpha\beta}^{ij}\simeq J+\Delta J_{\alpha\beta}^{ij},
\label{eq:jarl2}
\end{eqnarray}
 where the first term governs the $CP$-violating effects in the unitary limit
and the second term gives the contribution coming from the non-unitarity
effect:
\begin{eqnarray}
J&=&c_{12}c_{13}^2c_{23}s_{12}s_{13}s_{23}\sin\delta,\\
\Delta J_{\alpha\beta}^{ij} &\simeq & -\sum_{\gamma=e,\mu,\tau} {\rm Im}
\left(\eta_{\alpha\gamma} U_{\gamma i}U_{\beta j}U^*_{\alpha j}U^*_{\beta i}
+ \eta_{\beta\gamma}U_{\alpha i}U_{\gamma j}U^*_{\alpha j}U^*_{\beta i}
\right.\nonumber\\
&& \left. ~~~~~~~~~~~~~+
\eta^*_{\alpha\gamma}U_{\alpha i}U_{\beta j}U^*_{\gamma j}
U^*_{\beta i} +
\eta^*_{\beta \gamma}U_{\alpha i}U_{\beta j}U^*_{\alpha j}U^*_{\gamma i}\right)
\end{eqnarray}
Note that the unitary term $J$ vanishes if either
$s_{13}\to 0$ or $\delta\to 0$.
However, $\Delta J_{\alpha\beta}^{ij}$ depends on the off-diagonal
elements of $\eta$ (generally complex) and does not necessarily vanish even if
both $s_{13}$ and $\delta$ are zero; in fact, it might even dominate the
$CP$-violating effects in the leptonic sector.
 \item  The heavy neutrinos $N_i$ and $\tilde{N}_i$ entering the
charged-current sector can also mediate the rare lepton decays, $l_\alpha^-\to
 l_\beta^-\gamma$. Hence, unlike in the canonical seesaw model where this
contribution is suppressed by the light neutrino masses~\cite{depo},
 in this case it is constrained  mainly by the ratio $F=M_DM_N^{-1}$. The LFV
decays mediated by these heavy neutrinos have branching
ratios~\cite{pila}
\begin{eqnarray}
 {\rm BR}(l_\alpha\to l_\beta\gamma)\simeq
\frac{\alpha_W^3s_W^2m_{l_\alpha}^5}{256\pi^2M_W^4\Gamma_\alpha}
\left|\sum_{i=1}^6{\cal K}_{\alpha i}{\cal K}_{\beta i}^*I\left(\frac{m^2_{N_i}}{M^2_W}
\right)\right|^2
\label{eq:brlfv}
\end{eqnarray}
where $\Gamma_\alpha$ is the total decay width of $l_\alpha$ and the function $I(x)$ is defined by
\begin{eqnarray}
I(x)=-\frac{2x^3+5x^2-x}{4(1-x)^3}-\frac{3x^3\ln x}{2(1-x)^4}
\label{eq:ix}
\end{eqnarray}
 For degenerate RH neutrino masses, a reasonable assumption inspired by
resonant leptogenesis~\cite{blanchet}, the amplitude is proportional to
 $\left({\cal KK}^\dagger\right)_{\alpha\beta}\sim
\left(FF^\dagger\right)_{\alpha\beta}$, and hence, for sizeable $F$ and
TeV-scale RH sector, one could expect appreciable rates in the LFV channels.
On the
 other hand, in the conventional type I seesaw model, one has approximately
${\cal KK}^\dagger={\cal O}\left(m_\nu M_R^{-1}\right)$, and therefore,
the branching ratio,
${\rm BR}(l_\alpha\to l_\beta \gamma) \propto {\cal O}\left(m_\nu^2\right)$ is
strongly suppressed.
 \item The heavy neutrinos $N_i$ and $\tilde{N}_i$ also couple to the gauge
sector of the SM and can be produced on-shell, if kinematically accessible,
at hadron colliders via the gauge boson exchange diagrams. Due to their
pseudo-Dirac nature, the striking lepton number violating LHC signature of the
 fine-tuned type I and type III scenarios, namely
$pp\to l_\alpha^\pm l_\beta^\pm+$ jets, will be suppressed for
heavy Majorana states due to cancellation between
the graphs with internal lines of the $N$ and $\tilde{N}$ type which have
opposite $CP$-quantum numbers. However, the LFV processes are insensitive to
this effect and one can expect to get observable signals at the LHC. The most
distinctive signature would be the observation of LFV processes involving three
charged leptons in the final state plus missing energy, i.e.
$pp\to l_\alpha^\pm l_\beta^\pm l_\gamma^\mp \nu(\overline{\nu})+$ jets
~\cite{del}.
\end{enumerate}

 Thus we see that the phenomenology of the inverse seesaw mechanism depends
crucially on the mass ratio $F=M_DM_N^{-1}$. As noted earlier, we can
 choose the RH neutrino masses to be degenerate (with eigenvalue $m_N$), 
 inspired by resonant leptogenesis. So
we are left with a single mass parameter $m_N$, together with the Dirac
 mass matrix $M_D$ and the arbitrary small mass parameter $\mu$. In what
follows, we explicitly determine the form of $M_D$ in the context of
 a realistic supersymmetric $SO(10)$ model and then use the present
experimental bounds on the elements of the non-unitary parameter $|\eta|$ to
get a  lower bound on the RH neutrino mass scale $m_N$. Finally we fit the
observed LH neutrino mass and mixing parameters by the inverse seesaw formula
to determine the structure of $\mu$. We then study the phenomenological
consequences of our results.
\section{Embedding Inverse seesaw in realistic $SO(10)$ GUT}
As we have mentioned  earlier, in order to embed the inverse seesaw mechanism
into a supersymmetric $SO(10)$ theory, we have to break the $B-L$
symmetry by using a { \bf 16}~$\oplus~\overline{\bf 16}$ pair rather than a
{\bf 126}~$\oplus~\overline{\bf 126}$ pair of Higgs representation. In
this context, there are two symmetry breaking chains that are particularly
interesting:
\begin{itemize}
\item $SO(10)\stackrel{M_G}{\longrightarrow}
\mathbf 3_c \mathbf 2_L \mathbf 2_R \mathbf 1_{B-L}
\stackrel{M_R}{\longrightarrow}
\mathbf 3_c \mathbf 2_L \mathbf 1_Y ({\rm MSSM})
\stackrel{M_{\rm SUSY}}{\longrightarrow}
\mathbf 3_c \mathbf 2_L \mathbf 1_Y ({\rm SM})
\stackrel{M_Z}{\longrightarrow}
\mathbf 3_c \mathbf 1_Q$~\cite{desh}
\item $SO(10)\stackrel{M_G}{\longrightarrow}
\mathbf 3_c\mathbf 2_L\mathbf 2_R\mathbf 1_{B-L}
\stackrel{V_R}{\longrightarrow}
\mathbf 3_c\mathbf 2_L\mathbf 1_{I_{3R}}\mathbf 1_{B-L}
\stackrel{v_R}{\longrightarrow}
\mathbf 3_c\mathbf 2_L\mathbf 1_Y ({\rm MSSM})$\\
$~~~~~~~~~~~~~~~~~~~~~~~~~~~~~~~~~~~~~~~~~~~~~~~~~~~~~~~~~~
 \stackrel{M_{\rm SUSY}}{\longrightarrow}
\mathbf 3_c \mathbf 2_L \mathbf 1_Y ({\rm SM})
\stackrel{M_Z}{\longrightarrow} \mathbf 3_c \mathbf 1_Q$~\cite{ma05}
\end{itemize}
 where, as an example  of our notation, $\mathbf 3_c$ means $SU(3)_c$. In this
paper, we consider only the former (and simpler) case of $SO(10)$
 breaking chain. It was  shown in Ref.~\cite{desh} that it is possible to
obtain the gauge coupling unification in this model with a low-energy
(TeV-scale )  $SU(2)_R$ symmetry breaking  scale $M_R$. However, they
considered only one $\mathbf {10}_H$ Higgs field which contains only a single
bi-doublet (corresponding to the (1,2,2,0)  representation of
$\mathbf 3_c\mathbf 2_L\mathbf 2_R\mathbf 1_{B-L}$).  Getting a realistic
fermion mass  spectrum in this model is difficult (see, however, some recent
ideas~\cite{mimura} on how this could be done). Instead,
we consider a model with two $\mathbf {10}_H$ at the TeV scale. This
requires that we reexamine the unification issue with two Higgs bi-doublets.
We show that we not only obtain the gauge coupling unification at a
scale consistent with the proton decay bounds, but also successfully reproduce
the observed fermion masses and mixing.

 To study the running of the gauge couplings and the possibility of their
unification at a scale $M_G\sim 10^{16}$ GeV, we divide the whole energy
range $\left(M_Z,M_G\right)$ into three parts, according to the above
mentioned symmetry breaking chain:
\begin{itemize}
 \item First, we have the well known SM from the weak scale $M_Z$ to the
SUSY-breaking scale $M_{\rm SUSY}$ (which, for practical purposes, we assume
to be a little higher than $M_Z$);
 \item Then we have the MSSM from $M_{\rm SUSY}$ to the $B-L$ breaking scale
$M_R$ (which is assumed to be of order TeV, so that it is of
interest for colliders);
\item Finally, we have the Supersymmetric Left-Right (SUSYLR) model from $M_R$
to the unification scale $M_G$ (expected to be around $10^{16}$ GeV).
\end{itemize}

The running of the gauge couplings at one-loop level is determined by the RG
equation
\begin{eqnarray}
 \frac{d\alpha_i}{d\ln{\tilde{\mu}}} = \frac{b_i}{2\pi}\alpha_i^2,~~{\rm or,}~~
\alpha_i^{-1}(\tilde {\mu}_1) =
\alpha_i^{-1}(\tilde{\mu}_2)+\frac{b_i}{2\pi}\ln\left(\frac{\tilde{\mu}_2}
{\tilde{\mu}_1}\right)
\label{eq:gauge1}
\end{eqnarray}
where $\alpha_i\equiv \frac{g_i^2}{4\pi}$, $\tilde{\mu}$ is the energy scale,
and $b_i$'s are the coefficients of the one-loop $\beta$-functions. The SM and
MSSM $\beta$-functions are well known~\cite{martin}:
\begin{eqnarray}
b_i^{\rm SM}=\left(\frac{41}{10},-\frac{19}{6},-7\right),~~{\rm and}~~
b_i^{\rm MSSM}=\left(\frac{33}{5},1,-3\right),
\end{eqnarray}
where $i$ stands for $\mathbf 1_Y,~\mathbf 2_L$, and $\mathbf 3_c$
respectively. Before calculating the $\beta$-functions for the SUSYLR model,
let us first discuss the particle contents of this model.
\subsection{Particle Content of the SUSYLR Model}
 Here we consider only the doublet implementation of the SUSYLR model~
\cite{nick}, i.e. we use $SU(2)$ doublets of the ${\bf 16}_H$ Higgs field
 to break the $B-L$ symmetry.  In order to keep the model general, we allow
for an arbitrary number of these doublet fields, to be denoted by
 $n_L$ and $n_R$ respectively for $SU(2)_L$ and $SU(2)_R$ doublets.
Likewise we have $n_{10}$ bi-doublets of the ${\bf 10}_H$ Higgs field
 which, on acquiring VEVs, give masses to the fermions through Yukawa
couplings. We also allow for an arbitrary number $n_S$ of singlet fields
 $S^\alpha$. This is the minimal set of particles in a generic SUSYLR model.

However, it turns out that with this minimal set of particles, it is not
possible to obtain the gauge coupling unification at a scale higher than
$\sim 10^{15}$ GeV as required from current bounds on proton decay lifetime,
$\tau_p\gsim 10^{34}$ years~\cite{nishino}. As we have shown later in Section
4.2, unification is possible only after adding the contribution from the
color triplets $\left[\left(3,1,\frac{4}{3}\right)+{\rm c.c.}\right]$ which
come from the
$\mathbf {45}_H$ representation of Higgs at the unification scale $M_G$. It is
justified in Appendix A that it is indeed possible to have these color
triplets at TeV scale while all the other Higgs multiplets are still naturally
heavy at the GUT-scale. 
\begin{table}[h!]
\begin{center}
\begin{tabular}{||c|cccc||}\hline\hline
Multiplet & $SU(3)_c$ & $SU(2)_L$ & $SU(2)_R$ & $U(1)_{B-L}$\\
\hline
$Q$ & 3 & 2 & 1 & $+1/3$ \\
$Q^c$ & 3 & 1 & 2 & $-1/3$ \\
$L$ & 1 & 2 & 1 & $-1$ \\
$L^c$ & 1 & 1 & 2 & $+1$ \\
$\chi_p$ & 1 & 2 & 1 & $+1$ \\
$\chi^c_q$ & 1 & 1 & 2 & $-1$ \\
$\overline \chi_p$ & 1 & 2 & 1 & $-1$ \\
$\overline {\chi^c}_q$ & 1 & 1 & 2 & $+1$ \\
$\Phi_a$ & 1 & 2 & 2 & 0 \\
$S^\alpha$ & 1 & 1 & 1 & 0 \\
$\delta$ & 3 & 1 & 1 & $+4/3$ \\
$\overline{\delta}$ & $\bar{3}$ & 1 & 1 & $-4/3$\\
\hline\hline
\end{tabular}
\end{center}
 \caption{The representations of the particles under the
$\mathbf 3_c\mathbf 2_L\mathbf 2_R\mathbf 1_{B-L}$ gauge
group in the doublet SUSYLR model.
 Here $a=1,...,n_{10},~p=1,...,n_L,~q=1,...,n_R$ and
$\alpha=1,...,n_S$. The $B-L$ quantum numbers given
here are not GUT renormalized; to do so, we multiply a factor of
$\sqrt{\frac{3}{2}}$ (not $\sqrt{\frac{3}{8}}$ as mentioned in
Ref.~\cite{nick}).}
\end{table}

The particle content and their representations under the
$\mathbf 3_c\mathbf 2_L\mathbf 2_R\mathbf 1_{B-L}$ gauge
group are summarized in Table-1.
Following the notation of Ref.~\cite{nick}, the $SU(2)$ doublets and
bi-doublets are represented as
\[Q=\left(\begin{array}{c}
u \\ d \end{array}\right),~~
Q^c=\left(\begin{array}{c}
d^c \\ -u^c \end{array}\right),~~
\Phi_a=\left(\begin{array}{cc}
\phi^0_{a_d} & \phi^+_{a_u} \\
\phi^-_{a_d} & \phi^0_{a_u}
\end{array}\right) \]
Other doublet pairs can be written in a similar way as the $(Q,Q^c)$ pair.
The charges of the fields must obey the relation
\begin{equation}
Q=I_{3_L}+I_{3_R}+\frac{B-L}{2}
\end{equation}
\subsection{Gauge Coupling Unification}
The $\beta$-function for a general supersymmetric model is given by~
\cite{martin}
\begin{eqnarray}
b^{\rm SUSY}_N = 2n_g-3N+T(S_N)
\label{eq:beta}
\end{eqnarray}
 for $n_g$ generations of fermions, the gauge group $SU(N)$, and the complex
Higgs representation parametrized by $T(S_N)$. For $U(1)$ gauge group,
 $N=0$ in Eq.~(\ref{eq:beta}) and the gauge coupling is normalized as usual.
For the particle content given by Table-1, the Higgs contributions in our
SUSYLR model are explicitly given by
\begin{eqnarray}
T_{2L}=n_{10}+n_L,~T_{2R}=n_{10}+n_R,T_{3c}=1,~{\rm and}~
T_{B-L}=4+\frac{3}{2}\left(n_L+n_R\right)
\end{eqnarray}
Hence for three fermion generations, we find the $\beta$-functions for our
SUSYLR model to be
\begin{eqnarray}
b_i^{\rm SUSYLR}=\left(10+\frac{3}{2}n_L+\frac{3}{2}n_R,~n_{10}+n_L,~
n_{10}+n_R,~-2\right),
\label{eq:b1}
\end{eqnarray}
 where $i$ stands for $\mathbf 1_{B-L},~\mathbf 2_L,~\mathbf 2_R$ and
$\mathbf 3_c$ respectively. Using these $\beta$-functions, we can now obtain
the running of gauge couplings up to the scale $M_G$, knowing the initial
values at $\tilde{\mu}=M_Z$~\cite{pdg}
\begin{eqnarray}
\alpha_{1Y}(M_Z)&=&0.016829\pm 0.000017\nonumber\\
\alpha_{2L}(M_Z)&=&0.033493^{+0.000042}_{-0.000038}\nonumber\\
\alpha_{3c}(M_Z)&=&0.118\pm 0.003\nonumber
\end{eqnarray}
and the matching condition~\cite{book} at $\tilde{\mu}=M_R$ where the $U(1)_Y$-gauge
coupling gets merged into $SU(2)_R\times U(1)_{B-L}$:
\begin{eqnarray}
\alpha_{1Y}^{-1} (M_R) &=&
\frac{3}{5}\alpha_{2R}^{-1}(M_R)+\frac{2}{5}\alpha_{B-L}^{-1}(M_R)
\end{eqnarray}

 For numerical purposes, we assume $M_{\rm SUSY}=300$ GeV and $M_R=1$ TeV.
Also we take the number of Higgs bi-doublets,  $n_{10}=2$. However, the number
of Higgs doublets can be arbitrary and we vary these parameters to get the
unification. As shown in Figure-1, we
achieve the gauge unification for $n_L=0$ and $n_R=2$, with the
unification scale parameters
\begin{figure}[h!]
\centering
\includegraphics[width=10cm]{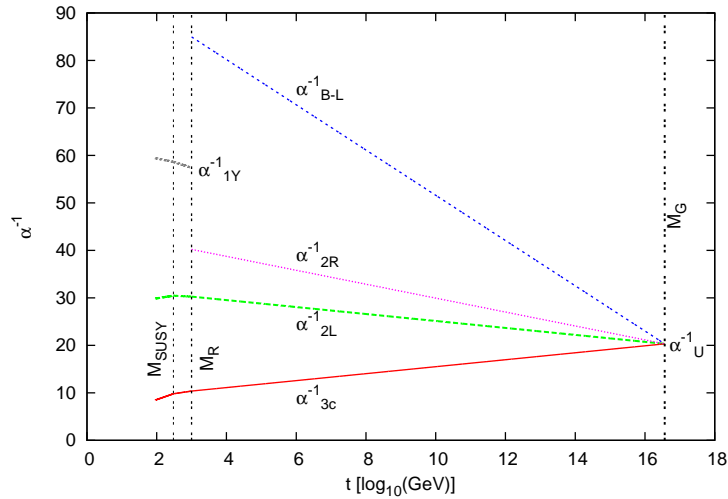}
\caption{Gauge coupling unification in the SUSYLR model. We have used
$n_{10}=2,~n_L=0,~n_R=2,~M_{\rm SUSY}=300$ GeV and $M_R=1$ TeV. As the 
running behavior is mostly controlled by the SUSYLR sector, the values of 
$M_{\rm SUSY}$ and $M_R$ can be relaxed a little bit, still preserving 
unification. However, it should be emphasized that the 
choice of the number of bi-doublets and 
doublets is the only possible choice consistent with both gauge coupling 
unification as well as 
realistic fermion masses. 
Increasing $n_{10}$ or $n_{L}$, or changing $n_R$ in either way, will spoil the 
unification, and as already noted, reducing $n_{10}$ will not give us a 
realistic fermion mass spectrum.} 
\end{figure}
\begin{eqnarray}
M_G\simeq 4\times 10^{16}~{\rm GeV},~~{\rm and}~~\alpha_U^{-1}(M_G)\simeq 20.3
\end{eqnarray}

Note the asymmetry between $n_L$ and $n_R$. We show in the appendix
that since the VEV of the ${\bf 45}_H$ Higgs breaks $D$-parity  
and decouples it from the $SU(2)_R$ breaking scale~\cite{dparity}, 
it is possible to have only the right-handed doublets and no left-handed ones
below the GUT scale. This leads to the asymmetry between $\alpha_{2L}$ 
and $\alpha_{2R}$, with $\frac{\alpha_{2L}}{\alpha_{2R}}\simeq 1.3$ 
in our case.

\section{RG evolution of the fermion masses and mixing}
 The RG evolution of the fermion masses and mixing have been extensively
studied for both the SM and the MSSM cases~\cite{das}, but not for the SUSYLR
 model, even though the analytical expressions for the Yukawa couplings have
already been derived in Ref.~\cite{nick}. Here we present a detailed
RG analysis in our SUSYLR model and obtain the quark and lepton masses and the
CKM matrix elements at the unification scale $M_G$.

The superpotential relevant for the RG evolution of the Yukawa couplings in
the SUSYLR model is given by~\cite{nick}
\begin{eqnarray}
W &\supset& ih_a Q^T \tau_2 \Phi_a Q^c + ih'_a L^T \tau_2 \Phi_a L^c +
i\lambda_{apq} \chi^T_p \tau_2 \Phi_a \chi^c_q +
i\bar\lambda_{apq} \overline{\chi}^T_p \tau_2 \Phi_a \overline{\chi^c}_q
\nonumber \\
&& + \mu^\Phi_{\alpha ab}S^\alpha {\rm Tr}\left(\Phi_a^T \tau_2 \Phi_b
\tau_2\right) +
i\mu^L_{\alpha p}S^\alpha L^T\tau_2 \chi_p + i\mu^{L^c}_{\alpha q}
S^\alpha L^{c^T}\tau_2 \chi^c_q
\label{eq:sup}
\end{eqnarray}
 where we have suppressed the generational and $SU(2)$ indices. Also we have
ignored all non-renormalizable terms in the superpotential as their
 contributions to the RGEs are suppressed by $M_R/M_G$. We note that the
superpotential given by Eq.~(\ref{eq:sup}) has two additional terms of the
 form $SL\chi$ and $SL^c\chi^c$ (as required by the inverse seesaw model) as
compared to that given in Ref.~\cite{nick}. Also note that since the
 $\delta,~\overline{\delta}$ fields do not couple to any of the matter fields,
they do not affect the renormalization group running except through their
effect on the color gauge coupling evolution.

  We have seen from the previous section that the gauge coupling unification
requires that we take two $SU(2)_R$ doublets and no $SU(2)_L$ doublet of
Higgs fields. Hence, dropping the $\chi,~\overline{\chi}$ terms from the
superpotential of Eq.~(\ref{eq:sup}), we have
\begin{eqnarray}
W &\supset& ih_a Q^T \tau_2 \Phi_a Q^c + ih'_a L^T \tau_2 \Phi_a L^c
+ i\mu^{L^c}_{\alpha q}S^\alpha L^{c^T}\tau_2 \chi^c_q
+ \mu^\Phi_{\alpha ab}S^\alpha {\rm Tr}\left(\Phi_a^T \tau_2 \Phi_b \tau_2\right)
\label{eq:modsup}
\end{eqnarray}
 where $a=1,2;~q=1,2$ and $\alpha=1,2,3$ corresponding to the two bi-doublets,
two RH doublets and three fermion singlets, respectively.

 The renormalization group equations (RGEs) for the Yukawa couplings $h_a$ and
$h'_a$ in Eq.~(\ref{eq:modsup}) are given by (with $t=\ln{\tilde{\mu}}$)
\begin{eqnarray}
 16\pi^2\frac{dh_a}{dt} &=& h_a\left[2h_b^\dagger
h_b-\frac{16}{3}g_3^2-3g_{2L}^2-3g_{2R}^2-\frac{1}{6}g_{B-L}^2\right]\nonumber\\
&&+ h_b\left[{\rm Tr}\left(3h_b^\dagger h_a+h_b'^{\dagger} h'_a\right)+2h_b^\dagger h_a
+4\left(\mu^{\Phi^\dagger}_\alpha \mu^\Phi_\alpha\right)_{ba}\right]
\label{eq:ha} \\
16\pi^2\frac{dh'_a}{dt} &=& h'_a\left[2h_b'^{\dagger} h'_b-3g_{2L}^2-3g_{2R}^2-\frac{3}{2}g_{B-L}^2\right]\nonumber\\
&&+ h'_b\left[{\rm Tr}\left(3h_b^\dagger h_a+h_b'^{\dagger} h'_a\right)+2h_b'^{\dagger} h'_a
+4\left(\mu^{\Phi^\dagger}_\alpha\mu^\Phi_\alpha\right)_{ba}
+\left(\mu^{L^c}\right)^\dagger_{\alpha q}\mu^{L^c}_{\alpha q}\delta_{ba}\right]
\label{eq:hap}
\end{eqnarray}
 where the repeated indices are summed over and $a,b=1,2;~q=1,2$; and
$\alpha=1,2,3~$ corresponding to the two Higgs
 bi-doublets, two $SU(2)_R$ doublets and three fermion singlets respectively.
Note that we have an additional contribution to the RGE of the
lepton Yukawa coupling $h_a'$ as compared to those given in
Ref.~\cite{nick} which comes from the $S\chi^c L^c$ term in the superpotential.
Note also the presence of the $h_b$ terms in the second line in
both the Yukawa runnings even for $a\neq b$, which are characteristic of
left-right models and are absent in the case of MSSM, arising from the Higgs
self energy effects.

 The fermion masses arise through the Yukawa couplings $h_a$ and $h'_a$ when
the two Higgs bi-doublets $\Phi_{1,2}$  acquire VEVs. In general, a linear
combination of $h_1$ and $h_2$ will give masses to the up-type quarks, and
 similarly different linear combinations for the other masses. The dynamics 
 of the superpotential can be chosen in such a way that the
bi-doublets acquire VEVs in the following simple manner:
\begin{eqnarray}
\langle \Phi_1\rangle = \frac{1}{\sqrt 2}\left(\begin{array}{cc} v_d & 0 \\ 0 & 0\end{array}\right),~~
\langle \Phi_2\rangle = \frac{1}{\sqrt 2}\left(\begin{array}{cc} 0 & 0 \\ 0 & v_u\end{array}\right)
\label{eq:bivev}
\end{eqnarray}
 and we identify the ratio $v_u/v_d\equiv \tan\beta$ (MSSM) with
$\sqrt{v_u^2+v_d^2}=246$ GeV. For numerical purposes,
we use $\tan\beta$ (MSSM)=10.
 To obtain the RGEs for the mass matrices, we choose the most frequently used
renormalization scheme~\cite{das}  where the Yukawa couplings and the Higgs
VEVs run separately. The RGEs for the Higgs VEVs are obtained from the gauge
and scalar self-energy contributions:
\begin{eqnarray}
    16\pi^2\frac{dv_u}{dt}&=&v_u\left[\frac{3}{2}g_{2L}^2+\frac{3}{2}g_{2R}^2-{\rm Tr}\left(3h_2^\dagger h_2+h_2'^{\dagger}h_2'\right)
-4\left(\mu^{\Phi^\dagger}_{\alpha}\mu^\Phi_\alpha\right)_{22}\right]
\label{eq:vu}\\
 16\pi^2\frac{dv_d}{dt}&=&v_d\left[\frac{3}{2}g_{2L}^2+\frac{3}{2}g_{2R}^2-{\rm Tr}\left(3h_1^\dagger
h_1+h_1'^{\dagger}h_1'\right)
-4\left(\mu^{\Phi^\dagger}_{\alpha}\mu^\Phi_\alpha\right)_{11}\right]
\label{eq:vd}
\end{eqnarray}
 Using Eqs.~(\ref{eq:ha}, \ref{eq:hap}) for $\dot{h}_a,~\dot{h}_a'$ and Eqs.
(\ref{eq:vu}, \ref{eq:vd}) for
 $\dot{v}_u,~\dot{v}_d$, we have derived the RGEs for the physical fermion
masses and the quark mixing in our SUSYLR
model in Appendix B. Using the initial values for the mass and mixing
parameters at $\tilde{\mu}=M_Z$~\cite{pdg}
\begin{eqnarray}
&&m_u(M_Z)=2.33^{+0.42}_{-0.45}~{\rm MeV},~~m_c(M_Z)=677^{+56}_{-61}~{\rm MeV},~~m_t(M_Z)=181\pm 13~{\rm GeV},\nonumber\\
 &&m_d(M_Z)=4.69^{+0.60}_{-0.66}~{\rm MeV},~~m_s(M_Z)=93.4^{+11.8}_{-13.0}~{\rm MeV},~~m_b(M_Z)=3.00\pm 0.11~{\rm
GeV},\nonumber\\
&&m_e(M_Z)=0.48684727\pm 0.00000014~{\rm MeV},\nonumber\\
&&m_\mu(M_Z)=102.75138\pm 0.00033~{\rm MeV},~~m_\tau(M_Z)=1.74669^{+0.00030}_{-0.00027}~{\rm GeV},\nonumber
\end{eqnarray}
 and with the quark-sector mixing parameters 
 $\theta_{12}=13.04^\circ\pm 0.05^\circ,~\theta_{13}=0.201^\circ\pm 0.011^\circ,~\theta_{23}=2.38^\circ \pm
0.06^\circ$ and $\delta_{13}=1.20\pm 0.08$,
\begin{eqnarray}
V_{\rm CKM}(M_Z)&=&\left(\begin{array}{ccc}
 c_{12}c_{13} & s_{12} c_{13} & s_{13}e^{-i\delta_{13}} \\ -s_{12}c_{23} - c_{12}s_{23}s_{13}e^{i\delta_{13}} &
 c_{12}c_{23} - s_{12}s_{23}s_{13}e^{i\delta_{13}} & s_{23}c_{13}\\ s_{12}s_{23} - c_{12}c_{23}s_{13}e^{i\delta_{13}}
& -c_{12}s_{23} - s_{12}c_{23}s_{13}e^{i\delta_{13}} & c_{23}c_{13}
\end{array}\right)\nonumber\\
&=&\left(\begin{array}{ccc}
0.9742 & 0.2256 & 0.0013-0.0033i\\
-0.2255-0.0001i & 0.9734 & 0.0415 \\
0.0081-0.0032i & -0.0407-0.0007i & 0.9991
\end{array}\right)\nonumber,
\end{eqnarray}
 and the SM and MSSM Yukawa RGEs~\cite{das} we numerically solve the SUSYLR
RGEs given in Appendix B to obtain
the running quark and lepton masses and the CKM matrix elements at the
unification scale $M_G$:
\begin{eqnarray}
&&m_u(M_G)=0.0017~{\rm GeV},~m_c(M_G)=0.1910~{\rm GeV},~m_t(M_G)=77.8035~{\rm GeV};\nonumber\\
&&m_d(M_G) = 0.0013~{\rm GeV},~m_s(M_G) =0.0263~{\rm GeV},~m_b(M_G) =1.7092~{\rm GeV} ;\nonumber\\
&&m_e(M_G) = 0.0004~{\rm GeV},m_\mu(M_G) =0.0911~{\rm GeV},~m_\tau(M_G)=1.7096~{\rm GeV}\nonumber\\
&&V_{\rm CKM}(M_G) = \left(\begin{array}{ccc}
0.9793 & 0.2023+0.0018i & 0.0005-0.0057i \\
-0.2023+0.0016i & 0.9791 & 0.0240 \\
0.0044-0.0056i & -0.0236-0.0013i & 0.9997
\label{eq:GUTmass}
\end{array}\right)
\label{eq:mgvql}
\end{eqnarray}
We also have a mild running for $\tan\beta$ with $\tan\beta(M_G)=7$ from
$\tan\beta(M_R)=10$.
\begin{figure}[h!]
\centering
\includegraphics[width=10cm]{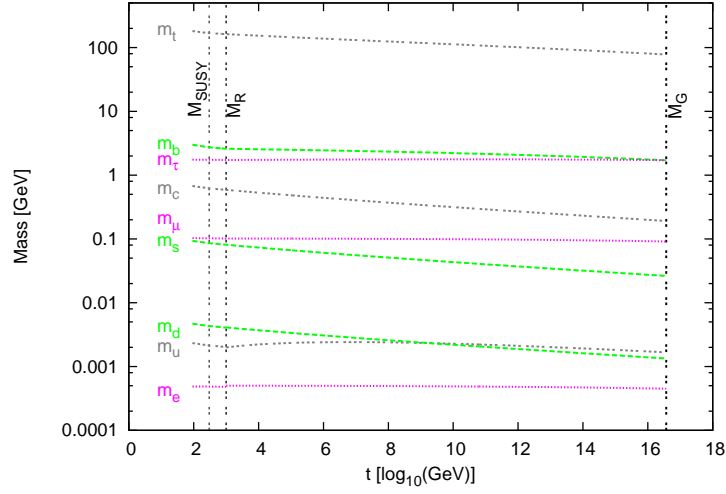}
\caption{Running of fermion masses to the GUT-scale in our model for
$M_{\rm SUSY}=300$ GeV and $M_R=1$ TeV. Note the $b-\tau$ unification which is 
a generic feature of GUT models.}
\end{figure}

\begin{figure}[h!]
\centering
\includegraphics[width=10cm]{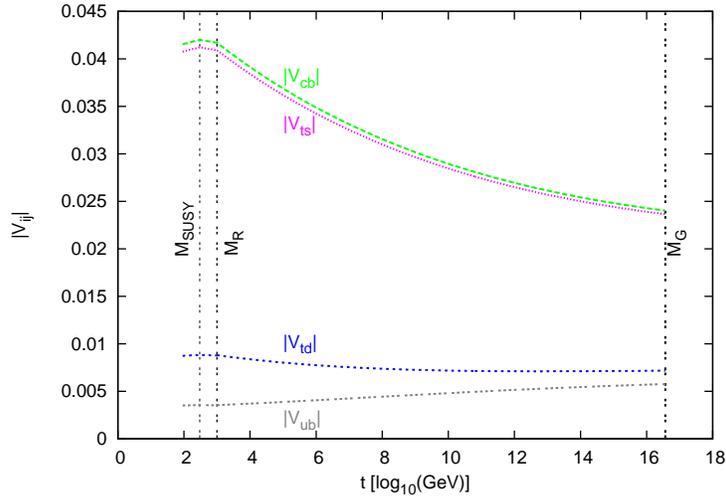}
\caption{Running of the CKM mixing elements involving third generation to the
GUT-scale in our model for
$M_{\rm SUSY}=300$ GeV and $M_R=1$ TeV. The running of other CKM elements, 
being small, is not shown here.}
\end{figure}

 Figure-2 shows the running of the quark and charged lepton masses up to the
unification scale $M_G$. Note that we are able to generate the 
fermion mass spectrum at the GUT scale with  
\begin{equation}
\frac{m^0_b}{m^0_\tau}\simeq 1,~\frac{m^0_\mu}{m^0_s}\simeq 3,~\frac{m^0_e}{m^0_d}\simeq \frac{1}{3}
\label{eq:pred}
\end{equation}
 Figure-3 shows the running of the CKM elements involving the third
generation. Note that in addition to the
 significant running for the third generation CKM elements
$V_{ub,~cb,~td,~ts}$, we have a relatively milder running
for the other elements as well [cf. Eq.~(\ref{eq:mgvql})], even in the 
third-generation dominance approximation. This is a characteristic of the 
Left-Right model, in contrast with 
the MSSM case where in the third generation dominance, the first and second
generation elements do not run at the one-loop level.
\section{The Dirac Mass for Neutrinos in a Specific $SO(10)$ Model}
 As discussed in Section 2, in order to implement the inverse seesaw
mechanism, we have to use the class of $SO(10)$
 models in which the $B-L$ subgroup is broken by a $\mathbf {16}_H\oplus
\overline{\mathbf {16}}_H$ pair. We also need
 at least two $\mathbf {10}_H$ and a $\mathbf {45}_H$ to have a realistic
fermion spectrum. With this minimum set of
 Higgs multiplets $\{\mathbf {10}_H,~\mathbf {16}_H,~\overline{\mathbf {16}}_H,
~\mathbf {45}_H\}$, several $SO(10)$
 models have been constructed~\cite{models}. All these models require 
 various dimension-5 operators to get right fermion masses: in principle, they are also present in our model. However, most of them e.g. $\frac{h_{ij}}{M}\mathbf {16}_i\mathbf {16}_j \mathbf {16}_H\mathbf {16}_H$, 
are suppressed by the factor $\frac{M_R}{M_{\rm Pl}}\sim 10^{-15}$ 
as the $\mathbf {16}_H$ Higgs acquires  only TeV-scale VEV. The only other  
dimension-5 operator
that can make significant contribution to fermion masses is  
$\frac{h'_{ij}}{M}\mathbf {16}_i\mathbf {16}_j\mathbf {10}_H\mathbf {45}_H$; 
we assume its effects to be small in our model and keep the dimension six operator
\begin{equation}
\frac{f_{ij}}{M^2}\mathbf {16}_i\mathbf {16}_j\mathbf {10}_H\mathbf {45}_H\mathbf {45}_H
\label{eq:45}
\end{equation}
  This operator is suppressed only by
  $\left(\frac{M_G}{M_{\rm Pl}}\right)^2\sim 10^{-4}$ as the $\mathbf {45}_H$ acquires a
VEV at the scale $M_G$ and plays an important role in the fermion mass fitting given below. 

The fermion mass
 splitting is obtained by the completely antisymmetric combination of the
operator given by the
expression~(\ref{eq:45}), i.e. in the notation of Ref.~\cite{sakita}
\begin{equation}
\langle \psi^*_+|B[\Gamma_i\Gamma_j\Gamma_k\Gamma_l\Gamma_m]A_{ij}A_{kl}\Phi_m|\psi_+\rangle
\end{equation}
 with $B=\displaystyle{\prod_{\mu={\rm odd}}}\Gamma_\mu$ and $[...]$ denoting
the completely antisymmetric  combination. Here $\Phi$ and $A$ denote the
$\mathbf {10}_H$ and $\mathbf {45}_H$ fields respectively. When the
following VEVs are non-zero:
\begin{equation}
\langle \Phi_{9,10}\rangle\neq 0,~~~\langle A_{12,34,56}\rangle\neq 0,
\end{equation}
 this antisymmetric combination acts as an effective $\mathbf {126}_H$
operator which gives the mass relation
 $m_e=-3m_d$ and $m_\nu=-3m_u$ due to the VEVs $\langle A_{ij} \rangle$,
while $m_u$ and $m_d$ are split in the usual
manner by the two $\mathbf {10}_H$ VEVs,  $\langle \phi_{9,10}\rangle$. To
obtain a realistic fermion mass spectrum, we construct the following model
using the  Higgs multiplets $\{\mathbf {10}_H,~\mathbf {45}_H,
~\mathbf {54}_H\}$. The $SO(10)$ symmetry breaking
 to $\mathbf 3_c\mathbf 2_L\mathbf 2_R\mathbf 1_{B-L}$ is obtained by a
combination of the $\mathbf {45}_H$ and
$\mathbf {54}_H$, with the following VEVs in an $SU(5)$ basis:
\begin{eqnarray}
\langle \mathbf {45}\rangle &\propto& {\rm diag}(a,a,a,0,0),\nonumber\\
\langle \mathbf {54}\rangle &\propto& {\rm diag}(2a,2a,2a,2a,2a,2a,-3a,-3a,-3a,-3a)
\end{eqnarray}
In this model, the fermion mass matrices at the GUT-scale have the following
form:
\begin{eqnarray}
M_u=\tilde{h}_u+\tilde{f},~~M_d=\tilde{h}_d+\tilde{f},~~M_e=\tilde{h}_d-3\tilde{f},~~M_D=\tilde{h}_u-3\tilde{f}
\label{eq:fit}
\end{eqnarray}
where the $h_{u,d}$ matrices come from the usual Yukawa terms $h_{ij}
\mathbf {16}_i\mathbf {16}_j\mathbf {10}_H(\mathbf {10}'_H)$ and the $f$
matrix comes from the $\mathbf {45}_H$ contribution given by the expression~
(\ref{eq:45}), where we have assumed the same coupling for both 
the ${\bf 10}_H$ fields. 
The tilde denotes the normalized couplings with mass dimensions
where
the VEVs have been absorbed. We know the nine eigenvalues of the quark and
charged lepton mass matrices at the scale $M_G$ from our RG analysis in
Section 5; however, we have 18 unknowns (for 3 hermitian matrices) to fit into
Eq.~(\ref{eq:fit}). Hence a unique fit is not possible; we just give here one
sample fit that is consistent with all the masses and mixing at the GUT scale
obtained from the RGEs.

We work in a basis in which the charged lepton mass matrix is diagonal, i.e.
\begin{eqnarray}
M_e=\left(\begin{array}{ccc}
m_e(M_G) & 0 & 0\\
0 & m_\mu(M_G) & 0\\
0 & 0 & m_\tau(M_G)\end{array}\right)
=\left(\begin{array}{ccc}
0.0004 & 0 & 0\\
0 & 0.0911 & 0\\
0 & 0 & 1.7096\end{array}\right)~{\rm GeV}\nonumber
\end{eqnarray}
This immediately implies from Eq.~(\ref{eq:fit}) that
\begin{equation}
\tilde{h}_{d,ij}=3\tilde{f}_{ij},~~~~~~~~~~~\forall~ i\neq j
\label{eq:foff}
\end{equation}
 For simplicity, let us choose the $\tilde{f}$-matrix to be diagonal. Then
Eq.~(\ref{eq:foff}) implies that $\tilde{h}_d$ is also a diagonal
matrix. We also have the following relations:
\begin{eqnarray}
\tilde{h}_{d,\alpha\alpha}+\tilde{f}_{\alpha\alpha}=m_\alpha,~~
\tilde{h}_{d,\beta\beta}-3\tilde{f}_{\beta\beta}=m_\beta
\label{eq:hds}
\end{eqnarray}
 where $m_\alpha=(m_d,~m_s,~m_b)$ are the eigenvalues of $M_d$ and
$m_\beta=(m_e,~m_\mu,~m_\tau)$ the eigenvalues of $M_e$.
These six equations (\ref{eq:hds}) now fix the $h_d$ and $f$ matrices
completely:
\begin{eqnarray}
\tilde{f}&=&\left(\begin{array}{ccc}
\frac{1}{4}\left(m_d-m_e\right) & 0 & 0\\
0 & \frac{1}{4}\left(m_s-m_\mu\right) & 0\\
0 & 0 & \frac{1}{4}\left(m_b-m_\tau\right)
\end{array}\right)\nonumber\\
& =& \left(\begin{array}{ccc}
2.25\times 10^{-4} & 0 & 0\\
0 & -0.0162 & 0 \\
0 & 0 & -0.0001
\end{array}\right)~{\rm GeV},\nonumber\\
\tilde{h}_d&=& \left(\begin{array}{ccc}
\frac{1}{4}\left(3m_d+m_e\right) & 0 & 0\\
0 & \frac{1}{4}\left(3m_s+m_\mu\right) & 0\\
0 & 0 & \frac{1}{4}\left(3m_b+m_\tau\right)
\end{array}\right)\nonumber\\
& =& \left(\begin{array}{ccc}
0.0011 & 0 & 0\\
0 & 0.0425 & 0 \\
0 & 0 & 1.7093
\end{array}\right)~{\rm GeV}
\end{eqnarray}
The $\tilde{h}_u$ matrix can now be determined by fitting to $M_u$ which, in
this basis, is given by
\begin{eqnarray}
M_u&=&V_{\rm CKM}M_u^{\rm diag}V_{\rm CKM}^\dagger\nonumber\\
&=& \left(\begin{array}{ccc}
0.0120 & 0.0384-0.0103i & 0.038-0.4433i\\
0.0384+0.0103i & 0.2280 & 1.8623+0.0002i\\
0.038+0.4433i & 1.8623-0.0002i & 77.7569
\end{array}\right)~{\rm GeV}
\end{eqnarray}
Then from Eq.~(\ref{eq:fit}) the $\tilde{h}_u$ matrix is given by
\begin{eqnarray}
\tilde{h}_u=\left(\begin{array}{ccc}
0.0118 & 0.0384-0.0103i & 0.038-0.4433i\\
0.0384+0.0103i & 0.2442 & 1.8623+0.0002i\\
0.038+0.4433i & 1.8623-0.0002i & 77.757
\end{array}\right)~{\rm GeV}
\end{eqnarray}
Hence the Dirac neutrino mass matrix is given by
\begin{eqnarray}
M_D=\left(\begin{array}{ccc}
0.0111 & 0.0384-0.0103i & 0.038-0.4433i\\
0.0384+0.0103i & 0.2928 & 1.8623+0.0002i\\
0.038+0.4433i & 1.8623-0.0002i & 77.7573
\end{array}\right)~{\rm GeV}
\label{eq:mnuso10}
\end{eqnarray}
It may be noted here that even though the specific form of the Dirac neutrino 
mass matrix may depend on the choice of the particular basis we have chosen, 
the individual values of the matrix elements are more or less fixed by the 
up-type quark mass values, due to the mass relation (\ref{eq:fit}), and hence, 
do not depend on the basis so much. Therefore, all the predictions of 
the model that follow from the form of $M_D$ given by Eq.~(\ref{eq:mnuso10}) 
will be independent of the initial choice of our basis, upto a few \%. 
\section{Non-unitarity effects in the lepton mixing matrix}
In this section we obtain the non-unitarity parameter $\eta$ using the
structure of the Dirac neutrino mass matrix obtained in Eq.~(\ref{eq:mnuso10})
and discuss the phenomenological consequences of our results.
\subsection{Bounds on $|\eta|$}
As discussed in Section 3, the non-unitarity parameter is given by
\begin{eqnarray}
\eta \simeq \frac{1}{2}FF^\dagger~~{\rm with}~F=M_DM_N^{-1}
\end{eqnarray}
For simplicity, choosing $M_N$ to be diagonal, and motivated by resonant
leptogenesis, assuming degenerate eigenvalues for $M_N$ equal to $m_N$,
we have
\begin{eqnarray}
\eta \simeq \frac{1}{2m_N^2}M_D M_D^\dagger
\end{eqnarray}
With the form of $M_D$ derived in the last section after
extrapolation to the weak scale, we can readily calculate the
elements of $\eta$:
\begin{eqnarray}
\eta \simeq \frac{1~{\rm GeV}^2}{m_N^2}\left(\begin{array}{ccc}
0.1 & 0.0412-0.4144i & 1.5134-17.247i\\
0.0412+0.4144i & 1.78 & 72.6794-0.0005i\\
1.5134+17.247i & 72.6794+0.0005i & 3024.93
\end{array}\right)
\label{eq:eta2}
\end{eqnarray}
This is to be compared with the present bounds on $|\eta_{ij}|$ (at the 90\%
C.L.) ~\cite{antusch}:
\begin{eqnarray}
|\eta| < \left(\begin{array}{ccc}
2.0\times 10^{-3} & 3.5\times 10^{-5} & 8.0\times 10^{-3} \\
3.5\times 10^{-5} & 8.0\times 10^{-4} & 5.1\times 10^{-3}\\
8.0\times 10^{-3} & 5.1\times 10^{-3} & 2.7\times 10^{-3}
\end{array}\right)
\end{eqnarray}
This gives a lower bound on the mass of the RH neutrino:
\begin{equation}
m_N \gsim 1.06~{\rm TeV},
\end{equation}
which should be kinematically accessible at the LHC to be produced
on-shell. Note that the right handed neutrinos are pseudo-Dirac
fermions in our model (with small Majorana component) which is
distinct from the type I seesaw models where they are pure
Majorana. As a result the like sign dilepton final states which
are the ``smoking gun'' collider signals of type I seesaw are
suppressed in our model; however, the trilepton signals can be
used in this case for testing these models~\cite{del}.

  With this lower bound on $m_N$, we get the following
improved bounds on $|\eta_{\alpha\beta}|$:
\begin{eqnarray}
&&|\eta_{ee}|<8.9\times 10^{-8},~~|\eta_{e\mu}|<3.7\times 10^{-7},~~|\eta_{e\tau}|<1.5\times 10^{-5},\nonumber\\
&&|\eta_{\mu\mu}|<1.6\times 10^{-6},~~|\eta_{\mu\tau}|<6.5\times 10^{-5}
\end{eqnarray}
At least one of these bounds, namely $|\eta_{e\mu}|$, is reachable
at future neutrino factories from the improved branching ratio of
$\mu\to e\gamma$ down to $10^{-18}$~\cite{schaaf}. Similar
sensitivities are also reachable in the PRISM/PRIME
project~\cite{prime}. We note that relaxing the condition of
degenerate RH neutrinos but fitting the neutrino masses affects
the values of $\eta_{\alpha\beta}$; we present these results in
Table-2. It appears that $|\eta_{e\mu}|$ values are all accessible to the
future $\mu\to e+\gamma$ searches; The largest value of
$|\eta_{\mu\tau}|$ in this table may also be accessible to
neutrino oscillation experiments, preferably with short baseline 
($L\lsim$ 100 km).
\begin{table}[h!]
\begin{center}
\begin{tabular}{||ccc|ccc||}\hline\hline
 $m_{N_1}$ & $m_{N_2}$ & $m_{N_3}$ & $|\eta_{e\mu}|$ & $|\eta_{e\tau}|$
& $|\eta_{\mu\tau}|$ \\ \hline\hline 1100 & 1100 & 1100 &
$3.7\times
10^{-7}$ & $1.5\times 10^{-5}$ & $6.5 \times 10^{-5}$ \\
100 & 100 & 1100 & $7.9\times 10^{-7}$ & $1.6\times 10^{-5}$ & $8.9\times 10^{-5}$\\
50 & 50 & 1200 & $2.5\times 10^{-6}$ &
$2.2\times 10^{-5}$ & $1.6\times 10^{-4}$ \\
30 & 30 & 2100 &
$6.7\times 10^{-6}$ & $ 4.4\times 10^{-5}$ & $3.2\times 10^{-4}$\\
\hline\hline
\end{tabular}
\end{center}
\caption{predictionss for the non-unitarity parameter
$|\eta_{\alpha\beta}|$ for the above choice of parameters in the
model including RH neutrino masses (given in GeVs).}
\end{table}
\subsection{Fitting the Neutrino Oscillation Data}
The structure of the small mass parameter $\mu$ can be obtained using the
inverse seesaw formula, Eq.~(\ref{eq:neutrino}):
\begin{eqnarray}
\mu = F^{-1}m_\nu\left(F^T\right)^{-1}
\label{eq:mu}
\end{eqnarray}
where $m_\nu$ is diagonalized by the new PMNS matrix ${\cal N}=(1-\eta)U$
instead of $U$ in Eq.~(\ref{eq:oldpmns}):
\begin{eqnarray}
m_\nu={\cal N}\widehat{m}_\nu {\cal N}^T
\end{eqnarray}
The form of $U$ is obtained from the standard PMNS parametrization using the
$2\sigma$ results from neutrino oscillation data~\cite{fogli}
\begin{eqnarray}
\Delta m^2_\odot &=& 7.67\left(1^{+0.044}_{-0.047}\right)\times 10^{-5}~{\rm eV}^2,\nonumber\\
\Delta m^2_{\rm atm} &=& 2.39\left(1^{+0.113}_{-0.084}\right)\times 10^{-3}~{\rm eV}^2,\nonumber\\
\sin^2\theta_{12} &=& 0.312\left(1^{+0.128}_{-0.109}\right),\nonumber\\
\sin^2\theta_{23} &=& 0.466\left(1^{+0.292}_{-0.215}\right),
\label{eq:ndata}
\end{eqnarray}
Here we assume $\theta_{13}=0$.
 Now using the form of $\eta$ obtained in
Eq.~(\ref{eq:eta2}) and taking $m_N=1.1$ TeV for its lower bound value, we
get the new PMNS matrix ${\cal N}=(1-\eta)U$.
For illustration, let us assume normal hierarchy for neutrino masses with
$m_1=10^{-3}$ eV. Then we obtain from Eq.~(\ref{eq:mu})
\begin{eqnarray}
\mu\simeq \left(\begin{array}{ccc}
-1.5934 + 0.0283i & 0.2244 - 0.0063i & -0.0044 + 0.0092i\\
0.2244 - 0.0063i & -0.0322 + 0.0012i & 0.0006 - 0.0013i\\
-0.0044 + 0.0092i & 0.0006 - 0.0013i & 4.0\times 10^{-5}+ 5.1\times 10^{-5}i
\end{array}\right)~{\rm GeV}
\end{eqnarray}
\subsection{$CP$-violation effects}
The $CP$-violation effects due to non-unitarity are measured by the Jarlskog
invariant $\Delta J_{\alpha\beta}^{ij}$ given by Eq.~(\ref{eq:jarl2}). Note
that $\Delta_{\alpha\beta}^{ij}$ is non-zero in our case as $\eta$ is a
complex matrix (the phases arising from the Dirac neutrino sector).
Using the values of $\theta_{ij}$ obtained from neutrino oscillation data
given by
Eqs.~(\ref{eq:ndata}) and the structure of $\eta$ determined in
Eq.~(\ref{eq:eta2}) with $m_N=1.1$ TeV, we obtain the following values for
$\Delta J_{\alpha\beta}^{ij}$:
\begin{eqnarray}
\Delta J_{e\mu}^{12} &\simeq & -2.4\times 10^{-6},\\
\Delta J_{e\mu}^{23} &\simeq & -2.7\times 10^{-6},\\
\Delta J_{\mu\tau}^{23} &\simeq & 2.7\times 10^{-6},\\
\Delta J_{\mu\tau}^{31} &\simeq & 2.7\times 10^{-6},\\
\Delta J_{\tau e}^{12} &\simeq & 7.1\times 10^{-6}
\end{eqnarray}
and $\Delta J_{e\mu}^{23} = \Delta J_{e\mu}^{31} = -\Delta J_{\mu\tau}^{12} = \Delta J_{\tau e}^{23} = \Delta J_{\tau e}^{31}$.
Note that these values are just one order of magnitude smaller than the
quark sector value, $J_{\rm CKM}=\left(3.05^{+0.19}_{-0.20}\right)
\times 10^{-5}$~\cite{pdg}, and can be the dominant source of $CP$-violation
in the leptonic sector for vanishing $\theta_{13}$, thus leading to 
distinctive $CP$-violating effects in neutrino oscillations~\cite{cpv1,cpv2}. 
For instance, the transition probability for the ``golden channel'' $\nu_\mu\to 
\nu_\tau$ with non-unitarity effects is given by~\cite{cpv1}
\begin{eqnarray}
	P_{\mu\tau} \simeq 4|\eta_{\mu\tau}|^2+4s_{23}^2c_{23}^2\sin^2
	{\left(\frac{\Delta m_{31}^2L}{4E}\right)}-4|\eta_{\mu\tau}|
	\sin{\delta_{\mu\tau}}s_{23}c_{23}\sin{\left(\frac{\Delta m_{31}^2L}{
	2E}\right)}
	\label{eq:pmutau}
\end{eqnarray}
where the last term is $CP$-odd due to the phase $\delta_{\mu\tau}$ of the 
element $\eta_{\mu\tau}$ which, in our model, is $\sim 7\times 10^{-6}$ 
[cf. Eq.~(\ref{eq:eta2})]. Hence, the $CP$-violating effects should be 
pronounced for long-baseline neutrino factories.  

\subsection{LFV decay rates}
Lepton flavor violating decays such as  $\mu\to e\gamma,~\tau\to
e\gamma$ and $\tau\to \mu\gamma$ are often a signature of seesaw
models for neutrino masses. In this model, they can arise from the
non-unitarity effects and can be obtained using
Eq.~(\ref{eq:brlfv}) which, for degenerate RH neutrinos, becomes :
\begin{eqnarray}
 {\rm BR}(l_\alpha\to l_\beta\gamma)\simeq
\frac{\alpha_W^3s_W^2m_{l_\alpha}^5}{256\pi^2M_W^4\Gamma_\alpha}
\left|\left({\cal K}{\cal K}^\dagger\right)_{\alpha\beta}I\left(\frac{m^2_{N}}{M^2_W}
\right)\right|^2,
\label{eq:brlfv2}
\end{eqnarray}
with ${\cal K}=V_{3\times 6}$ and $I(x)$ defined in Eq.~(\ref{eq:ix}).
Now that we know all the three $3\times 3$ mass matrices entering the
inverse seesaw formula
given by Eq.~(\ref{eq:big}), we can easily determine the
structure of the full unitary matrix $V$ by diagonalizing the $9\times 9$
neutrino mass matrix ${\cal M}_\nu$, and hence, obtain $V_{3\times 6}$.

The total decay width $\Gamma_\alpha$ entering Eq.~(\ref{eq:brlfv2}) is given
by $\hbar/\tau_\alpha$ where the mean life for $\mu$ and $\tau$ are,
respectively~\cite{pdg}
\begin{eqnarray}
\tau_\mu &=& (2.197019\pm 0.000021)\times 10^{-6}~{\rm sec.},\nonumber\\
\tau_\tau &=& (290.6\pm 1.0)\times 10^{-15}~{\rm sec.}\nonumber
\end{eqnarray}
Using these values, we obtain the following branching ratios for the rare LFV
decays
\begin{eqnarray}
{\rm BR}(\mu\to e\gamma) &\simeq & 3.5\times 10^{-16},\\
{\rm BR}(\tau\to e\gamma) &\simeq & 1.1\times 10^{-13},\\
{\rm BR}(\tau\to \mu\gamma) &\simeq & 2.0\times 10^{-12}
\end{eqnarray}
We have estimated the contribution to $\mu\to e+\gamma$ branching
ratio from the off diagonal Dirac Yukawa coupling contribution to
slepton masses and find that for universal scalar mass of 500 GeV
and tan $\beta\simeq 5$, it is comparable to this value or less.
Such values for $\mu\to e\gamma$ branching ratio are accessible to
future experiments~\cite{schaaf,prime} capable of reaching
sensitivities down to $10^{-18}$. They can be used to test the
model. 

In our model we assume that squark and slepton masses are above a
TeV so that their contribution to the flavor changing neutral
current effects are negligible. 
The predictions for $\mu\to 3 e$ and $\mu\to e$ conversion~\cite{pilalfv} 
for a TeV-scale slepton mass, as in our model, 
are much smaller than what can be probed in planned experiments.
\section{Summary}
In conclusion, we have presented a TeV scale realistic inverse
seesaw scenario that arises from a supersymmetric $SO(10)$ model
consistent with gauge coupling unification and fermion mass
spectrum. This required us to carry out an extrapolation of quark
masses and mixing to the GUT scale with a TeV scale SUSYLR rather
than MSSM. This appears to be the first time that such an
extrapolation is carried out. Implementation of inverse seesaw
within the $SO(10)$ helps to reduce the number of parameters
making the model predictive. We present our expectations for the
non-unitarity of the PMNS leptonic mixing matrix with the choice
of parameters and its other phenomenological consequences. The
heavy RH neutrinos which are pseudo-Dirac fermions have TeV scale
mass and can be produced in colliders, thus giving rise to
distinctive signatures. We also give our predictions (with our
choice of parameters) for the non-unitarity contribution to the
branching ratios for the rare LFV decays of muons and taus. The
model can also be tested by the production of $W_R$ and $Z'$
bosons which are at the TeV scale. Of these, the branching ratio
$\mu\to e+\gamma$ could be testable in future experiments. Some of
the elements of the non-unitarity matrix $|\eta|$ predicted by our
model may be accessible to the next generation neutrino factories
too.

\section*{Acknowledgments}
This work is supported by the NSF under grant No. PHY-0652363. B.D. would like
to thank Aleksandr Azatov for discussions.
\section*{Appendix A: Masses of the $SO(10)$ Higgs multiplets}
As discussed in Section 4.2, we  obtain the gauge coupling unification at an
acceptable scale only after including the contribution from the color triplets
$\delta\left(3,1,1,+\frac43\right),~\overline\delta\left(\bar{3},1,1,-\frac43
\right)$. This pair of Higgs
fields is contained in the $\mathbf {45}$ representation
of Higgs in a generic $SO(10)$ model. However, in principle, there could be
other light gauge multiplets of $\mathbf {45}$ and/or
$\mathbf {54}$ that might contribute to the gauge coupling running as well.
Here we argue that in a generic $SO(10)$ model with only
$\mathbf {45}_H$ and $\mathbf {54}_H$  representations of Higgs (apart from
the essential $\mathbf {10}_H$ and $\mathbf {16}_H$), it is possible to have
only the $\delta$'s as light states (TeV scale) whereas all the other states
are very heavy at GUT scale, and hence, do not contribute to the RG running.
It turns out that we need to have at least two $\mathbf {45}_H$'s in
our model in order to have these light color triplets.

The most general Higgs superpotential with two $\mathbf A\equiv\mathbf {45}$'s
and a $\mathbf E\equiv\mathbf {54}$ Higgs
fields is given by
\begin{eqnarray}
W_H&=& \frac{1}{2}m_1\mathbf A^2+ \frac{1}{2}m'_1{\mathbf A'}^2+
\frac{1}{2}m_2\mathbf E^2+\lambda_1 \mathbf E^3+
\lambda_2 \mathbf E\mathbf A^2+\lambda'_2\mathbf E\mathbf A'^2
+\lambda_3\mathbf E\mathbf A\mathbf A'
\label{eq:higgssup}
\end{eqnarray}
where we have absorbed the $\mathbf A\mathbf A'$ term by a
redefinition of the fields. The Higgs fields $\mathbf A,~\mathbf A'$ and
$\mathbf E$ contain three
directions of singlets (with $\mathbf A$ and $\mathbf A'$ VEVs parallel)
under the SM subgroup $\mathbf 3_c\mathbf 2_L\mathbf 1_Y$~\cite{fuku}.
The corresponding VEVs are defined by
\begin{eqnarray}
\langle \mathbf A\rangle =\sum_{i=1}^2 A_i\widehat{A}_i,~~~~
\langle \mathbf A'\rangle =\sum_{i=1}^2 A'_i\widehat{A'}_i,~~~~
\langle \mathbf E \rangle =E\widehat{E}
\label{eq:unit}
\end{eqnarray}
where in the notation of Ref.~\cite{fuku}, the unit directions $\widehat{A}_i$
and $\widehat{E}$ in the $Y$-diagonal basis are given by
\begin{eqnarray}
\widehat{A}_1 &=& \widehat{A}_{(1,1,3)}^{(1,1,0)} = \frac{i}{2}[78+90]
=\widehat{A'}_1,\nonumber\\
\widehat{A}_2 &=& \widehat{A}_{(15,1,1)}^{(1,1,0)} = \frac{i}{\sqrt 6}[12+34+56]=\widehat{A'}_2,\nonumber\\
\widehat{E} &=& \widehat{E}_{(1,1,1)}^{(1,1,0)} = \frac{1}{\sqrt{60}}\left(-2\times [12+34+56]+3\times
[78+90]\right)
\label{eq:defunit}
\end{eqnarray}
where the upper and lower indices denote the $\mathbf 3_c\mathbf 2_L\mathbf 1_Y$ and $\mathbf 4_c\mathbf 2_L\mathbf 2_R$
quantum numbers respectively. The unit directions in Eq.~(\ref{eq:unit}) satisfy the orthonormality relations
\begin{eqnarray}
\widehat{A}_i\cdot \widehat{A}_j=\delta_{ij}~~~~{\rm and}~~\widehat{E}\cdot \widehat{E}=1
\label{eq:ortho}
\end{eqnarray}
The superpotential of Eq.~(\ref{eq:higgssup}) calculated at the VEVs in Eq.~(\ref{eq:unit}) is given by
\begin{eqnarray}
 \langle W_H\rangle &=& \frac{1}{2}m_1\langle A\rangle^2+
\frac{1}{2}m'_1\langle A'\rangle^2+
\frac{1}{2}m_2\langle E\rangle^2\nonumber\\
&&+\lambda_1\langle
E\rangle^3+\lambda_2 \langle E \rangle \langle A\rangle^2
+\lambda'_2 \langle E \rangle \langle A'\rangle^2
+\lambda_3 \langle E \rangle \langle A \rangle \langle A' \rangle
\nonumber\\
 &=&\frac{1}{2}m_1(A_1^2+A_2^2)
+\frac{1}{2} m'_1(A_1'^2+A_2'^2)
+\frac{1}{2}m_2E^2+\frac{\lambda_1}{2\sqrt{15}}E^3\nonumber\\
&&+
\frac{E}{2\sqrt{15}}\left[\lambda_2(3A_1^2-2A_2^2)
+\lambda'_2(3A_1'^2-2A_2'^2)+\lambda_3(3A_1A'_1-2A_2A'_2)\right]
\end{eqnarray}
 using the definitions in Eqs.~(\ref{eq:defunit}) and the orthonormality
relations given by Eqs.~(\ref{eq:ortho}). The VEVs are determined by the minimization of the
superpotential with respect to the fields:
\begin{eqnarray}
 \left\{\frac{\partial}{\partial A_1},~\frac{\partial}{\partial A_2},~
\frac{\partial}{\partial A'_1},~\frac{\partial}{\partial A'_2},~
\frac{\partial}{\partial E}\right\}\langle
W_H\rangle =0
\end{eqnarray}
This yields a set of five equations for $A_1,~A_2,~A'_1,~A'_2$ and $E$:
\begin{eqnarray}
0 &=& m_1A_1+\frac{3}{\sqrt{15}}\lambda_2EA_1
+\frac{3}{2\sqrt{15}}\lambda_3EA'_1,\nonumber\\
0 &=& m_1A_2-\frac{2}{\sqrt{15}}\lambda_2EA_2
-\frac{2}{2\sqrt{15}}\lambda_3EA'_2,\nonumber\\
0 &=& m'_1A'_1+\frac{3}{\sqrt{15}}\lambda'_2EA'_1
+\frac{3}{2\sqrt{15}}\lambda_3EA_1,\label{eq:1vev}\\
0 &=& m'_1A'_2-\frac{2}{\sqrt{15}}\lambda'_2EA'_2
-\frac{2}{2\sqrt{15}}\lambda_3EA_2,\nonumber\\
0 &=& m_2E+\frac{1}{2\sqrt{15}}\left[3\lambda_1E^2
+\lambda_2(3A_1^2-2A_2^2)+\lambda'_2(3A_1'^2-2A_2'^2)
+\lambda_3(3A_1A'_1-2A_2A'_2)\right]\nonumber
\label{eq:3vev}
\end{eqnarray}
As in our model, the $SO(10)$ symmetry is broken by the $\mathbf {45}$ and
$\mathbf {54}$ VEVs to
$\mathbf 3_c\mathbf 2_L\mathbf 2_R\mathbf 1_{B-L}$ gauge group at the scale
$M_G$,
we are interested in the $\mathbf 3_c\mathbf 2_L\mathbf 2_R\mathbf 1_{B-L}$
symmetry solutions~\cite{fuku}
\[A_1=A'_1=0,~~A_2\neq 0,~~A'_2\neq 0,~~E\neq 0\]
Hence it follows from Eqs.~(\ref{eq:1vev}) that
\begin{eqnarray}
 m_1 - \frac{2\lambda_2E}{\sqrt{15}}=\frac{\lambda_3E}{\sqrt{15}}
\frac{A'_2}{A_2},~~~
 m'_1 - \frac{2\lambda'_2E}{\sqrt{15}}=\frac{\lambda_3E}{\sqrt{15}}
\frac{A_2}{A'_2}
\label{eq:m12}
\end{eqnarray}

In order to study the mass matrices, it is convenient to decompose the Higgs
representations under the SM gauge group
$\mathbf 3_c\mathbf 2_L\mathbf 1_Y$. In Table-3 we present the explicit
decompositions of all the Higgs representations under the chain of subgroups
 \[\mathbf 4_c\mathbf 2_L\mathbf 2_R \supset \mathbf 3_c\mathbf 2_L\mathbf 2_R\mathbf 1_{B-L} \supset \mathbf
3_c\mathbf 2_L\mathbf 1_Y.\]
 Using the Clebsch-Gordan coefficients given in Ref.~\cite{fuku}, we obtain
the masses  of these multiplets as follows. The basis designating the columns
(c) of the mass matrices is given in the same way as in
 Table-3 while the rows (r) are designated by the corresponding complex
conjugated $\mathbf 3_c\mathbf 2_L\mathbf 1_Y$ multiplets.

First, we obtain the masses of the multiplet
$\left[\left(3,1,\frac{4}{3}\right)+{\rm c.c.}\right]$ in the basis
\begin{eqnarray}
&&{\rm c}~:~ \widehat {A}_{(15,1,1)}^{\left(3,1,\frac{4}{3}\right)},
~\widehat {A'}_{(15,1,1)}^{\left(3,1,\frac{4}{3}\right)};~~
{\rm r}~:~\widehat {A}_{(15,1,1)}^{\left(\bar{3},1,-\frac{4}{3}\right)},~
\widehat {A'}_{(15,1,1)}^{\left(\bar{3},1,-\frac{4}{3}\right)}
\nonumber\\
&&M_\delta^{\left(3,1,\frac43\right)}=\left(\begin{array}{cc}
m_1-\frac{2\lambda_2E}{\sqrt{15}} & -\frac{\lambda_3E}{\sqrt{15}}\\
-\frac{\lambda_3E}{\sqrt{15}} & m'_1-\frac{2\lambda'_2E}{\sqrt{15}}
\end{array}\right)=\frac{\lambda_3E}{\sqrt{15}}
\left(\begin{array}{cc}
\frac{A'_2}{A_2} & -1\\
-1 & \frac{A_2}{A'_2}
\end{array}\right)
\end{eqnarray}
using Eq.~(\ref{eq:m12}). It is obvious that ${\rm det}(M_\delta)=0$, and hence,
one of the two eigenvalues is zero while the other eigenvalue is given by
\begin{eqnarray}
{\rm Tr}(M_\delta)=\frac{\lambda_3E}{\sqrt{15}}
\left(\frac{A'_2}{A_2}+\frac{A_2}{A'_2}\right),
\label{eq:mdelta}
\end{eqnarray}
The zero eigenvalues (six in total) are easily identified as the longitudinal
Nambu-Goldstone modes as the $SU(4)_c$ gauge group breaks to
$SU(3)_c\times U(1)_{B-L}$ and they acquire mass of order $M_G$ by the usual
Higgs mechanism once the $\mathbf {45}_H$ gets VEV at the GUT scale.
We keep the other six
eigenvalues given by Eq.~(\ref{eq:mdelta}) at TeV scale by fine-tuning the
coupling $\lambda_3$. In what follows, we explicitly calculate the mass
eigenvalues for all the other multiplets given by Table-3 and show that it is
possible to have only the above six massive $\delta$'s at the TeV scale while
all the other states of $\mathbf {45}$ and $\mathbf {54}$ are heavy at the
GUT-scale.

We note that once we assume $\lambda_3$ to be small, the effect of
the second $\mathbf {45}_H$ multiplet becomes negligible and we can as well
drop the primed terms in the superpotential. For simplicity, we also assume
that $A_2=E\sim M_G$. Then the VEV conditions given by Eqs.~(\ref{eq:1vev})
yield
\begin{eqnarray}
m_1\simeq \frac{2\lambda_2E}{\sqrt{15}},~~
m_2\simeq \frac{E}{\sqrt{15}}\left(\lambda_2-\frac32\lambda_1\right)
\end{eqnarray}
We list below the mass eigenvalues for all the multiplets given in Table-3.
\begin{table}[h!]
\begin{center}
\begin{tabular}{||c|c|c|c||}\hline\hline
 $SO(10)$ & $\mathbf 4_c,\mathbf 2_L,\mathbf 2_R$ & $\mathbf 3_c,\mathbf 2_L,\mathbf 2_R,\mathbf 1_{B-L}$ & $\mathbf
3_c,\mathbf 2_L,\mathbf 1_Y$\\
\hline\hline
& (1,2,2) & (1,2,2,0) & $(1,2,\pm 1)$\\ \cline{2-4}
\bf {10} & (6,1,1) & $\left(3,1,1,-\frac23\right)$ &
$\left(3,1,-\frac23\right)$\\
& &$ \left(\bar 3,1,1,\frac23\right)$ & $\left(\bar 3,1,\frac23\right)$\\
\hline\hline
& (4,2,1) & $\left(3,2,1,\frac13\right)$ & $\left(3,2,\frac13\right)$\\
& & $(1,2,1,-1)$ & $(1,2,-1)$ \\ \cline{2-4}
\bf {16} &  & $\left(\bar 3,1,2,-\frac13\right)$ & $\left(\bar 3,1,\frac23\right)$\\
& $(\bar 4,1,2)$ & & $\left(\bar 3,1,-\frac43\right)$\\ \cline{3-4}
& & (1,1,2,1) & (1,1,2)\\
& & & (1,1,0) \\
\hline\hline
& & & (1,1,2) \\
& (1,1,3) & (1,1,3,0) & (1,1,0) \\
& & & $(1,1,-2)$ \\ \cline{2-4}
& (1,3,1) & (1,3,1,0) & (1,3,0) \\ \cline{2-4}
 & & $\left(3,2,2,-\frac{2}{3}\right)$ & $\left(3,2,\frac{1}{3}\right)$ \\
$\mathbf {45}$ & (6,2,2) & & $\left(3,2,-\frac{5}{3}\right)$ \\ \cline{3-4}
& & $\left(\bar{3},2,2,\frac{2}{3}\right)$ & $\left(\bar{3},2,\frac{5}{3}\right)$\\
& & & $\left(\bar{3},2,-\frac{1}{3}\right)$ \\ \cline{2-4}
& & (1,1,1,0) & (1,1,0)\\
& (15,1,1) & $\left(3,1,1,\frac{4}{3}\right)$ & $\left(3,1,\frac{4}{3}\right)$\\
& & $\left(\bar{3},1,1,-\frac{4}{3}\right)$ & $\left(\bar{3},1,-\frac{4}{3}\right)$\\
& & (8,1,1,0) & (8,1,0)\\ \hline\hline
& (1,1,1) & (1,1,1,0) & (1,1,0) \\ \cline{2-4}
& & & (1,3,2)\\
& (1,3,3) & (1,3,3,0) & (1,3,0)\\
& & & $(1,3,-2)$ \\ \cline{2-4}
& & $\left(3,2,2,-\frac{2}{3}\right)$ & $\left(3,2,\frac{1}{3}\right)$ \\
$\mathbf {54}$ & (6,2,2) & & $\left(3,2,-\frac{5}{3}\right)$\\ \cline{3-4}
& & $\left(\bar{3},2,2,\frac{2}{3}\right)$ & $\left(\bar{3},2,\frac{5}{3}\right)$\\
& & & $\left(\bar{3},2,-\frac{1}{3}\right)$ \\ \cline{2-4}
& & $\left(6,1,1,-\frac{4}{3}\right)$ & $\left(6,1,-\frac{4}{3}\right)$\\
& $(20',1,1)$ & $\left(\bar{6},1,1,\frac{4}{3}\right)$ & $\left(\bar{6},1,\frac{4}{3}\right)$\\
& & (8,1,1,0) & (8,1,0) \\
\hline\hline
\end{tabular}
 \caption{Decomposition of the {\bf 10},~{\bf 16},~{\bf 45} and
{\bf 54} Higgs representations under the chain of subgroups
 $\mathbf 4_c\mathbf 2_L\mathbf 2_R \supset \mathbf 3_c\mathbf 2_L\mathbf 2_R\mathbf 1_{B-L} \supset \mathbf
3_c\mathbf 2_L\mathbf 1_Y$.}
\end{center}
\end{table}
\begin{itemize}
\item (1,1,0) : We have three such states and the mass matrix is given by
\begin{eqnarray}
 &&{\rm c}~:~ \widehat {A}_{(1,1,3)}^{(1,1,0)},~\widehat {A}_{(15,1,1)}^{(1,1,0)},~\widehat
{E}_{(1,1,1)}^{(1,1,0)}; ~~{\rm r}~:~ \widehat {A}_{(1,1,3)}^{(1,1,0)},~\widehat {A}_{(15,1,1)}^{(1,1,0)},~\widehat
{E}_{(1,1,1)}^{(1,1,0)}\nonumber\\
&&\left(\begin{array}{ccc}
m_1+\frac{3\lambda_2E}{\sqrt{15}} & 0 & \frac{3\lambda_2A_1}{\sqrt{15}}\\
0 & m_1-\frac{2\lambda_2E}{\sqrt{15}} & -\frac{2\lambda_2A_2}{\sqrt{15}}\\
\frac{3\lambda_2A_1}{\sqrt{15}} & -\frac{2\lambda_2A_2}{\sqrt{15}} & m_2+\frac{3\lambda_1E}{\sqrt{15}}
\end{array}\right)=\frac{E}{\sqrt{15}}\left(\begin{array}{ccc}
5\lambda_2 & 0 & 0\\
0 & 0 & -2\lambda_2\\
0 & -2\lambda_2 & \lambda_2+\frac{3}{2}\lambda_1
\end{array}\right)\nonumber
\end{eqnarray}
So the mass eigenvalues are
\begin{eqnarray}
 M_1^{(1,1,0)} &=& \frac{5E}{\sqrt{15}}\lambda_2 \neq 0,\nonumber\\
 M_2^{(1,1,0)} &=&
 \frac{E}{2\sqrt{15}}\left[\left(\lambda_2+\frac{3}{2}\lambda_1\right)+
\sqrt{\left(\lambda_2+\frac{3}{2}\lambda_1\right)^2+16\lambda_2^2}\right] \neq 0,\nonumber\\
M_3^{(1,1,0)} &=& \frac{E}{2\sqrt{15}}\left[\left(\lambda_2+\frac{3}{2}\lambda_1\right)
-\sqrt{\left(\lambda_2+\frac{3}{2}\lambda_1\right)^2+16\lambda_2^2}\right] \neq 0
\end{eqnarray}
\item $[(1,1,2)+{\rm c.c.}]$ : There is only one such multiplet and its mass is
\begin{eqnarray}
&&{\rm c}~:~\widehat A_{(1,1,3)}^{(1,1,2)},~~{\rm r}~:~\widehat A_{(1,1,3)}^{(1,1,-2)}\nonumber\\
&& M^{(1,1,2)} = m_1+\frac{3}{\sqrt{15}}\lambda_2E = \frac{5E}{\sqrt{15}}\lambda_2\neq 0
\end{eqnarray}
\item  $\left[\left(3,2,-\frac{5}{3}\right)+{\rm c.c.}\right]$ : There are two such multiplets and the mass matrix is
\begin{eqnarray}
 &&{\rm c}~:~\widehat A_{(6,2,2)}^{\left(3,2,-\frac{5}{3}\right)},~\widehat
E_{(6,2,2)}^{\left(3,2,-\frac{5}{3}\right)};~~
{\rm r}~:~\widehat A_{(6,2,2)}^{\left(\bar{3},2,\frac{5}{3}\right)},~\widehat
E_{(6,2,2)}^{\left(\bar{3},2,\frac{5}{3}\right)}\nonumber\\
&&\left(\begin{array}{cc}
m_1+\frac{\lambda_2E}{2\sqrt{15}} & -\frac{\lambda_2A_1}{2}-\frac{\lambda_2A_2}{\sqrt 6}\\
-\frac{\lambda_2A_1}{2}-\frac{\lambda_2A_2}{\sqrt 6} & m_2+\frac{3\lambda_1E}{2\sqrt{15}}
\end{array}\right)= \frac{\lambda_2E}{\sqrt{15}}\left(\begin{array}{cc}
3 & -\sqrt{\frac{5}{2}}\\
-\sqrt{\frac{5}{2}} & 1
\end{array}\right)\nonumber
\end{eqnarray}
with the eigenvalues
\begin{eqnarray}
M_{1,2}^{\left(3,2,-\frac{5}{3}\right)} = \frac{E\lambda_2}{2\sqrt{15}}\left[4\pm \sqrt{14}\right] \neq 0
\end{eqnarray}
\item $\left[\left(3,2,\frac{1}{3}\right)+{\rm c.c.}\right]$ : There are two of them and the mass matrix is
\begin{eqnarray}
 &&{\rm c}~:~\widehat A_{(6,2,2)}^{\left(3,2,\frac{1}{3}\right)},~\widehat
E_{(6,2,2)}^{\left(3,2,\frac{1}{3}\right)};~~
{\rm r}~:~\widehat A_{(6,2,2)}^{\left(\bar{3},2,-\frac{1}{3}\right)},~\widehat
E_{(6,2,2)}^{\left(\bar{3},2,-\frac{1}{3}\right)}\nonumber\\
&&\left(\begin{array}{cc}
m_1+\frac{\lambda_2E}{2\sqrt{15}} & \frac{\lambda_2A_1}{2}-\frac{\lambda_2A_2}{\sqrt{6}}\\
\frac{\lambda_2A_1}{2}-\frac{\lambda_2A_2}{\sqrt{6}} & m_2+\frac{3\lambda_1E}{2\sqrt{15}}
\end{array}\right)=\frac{\lambda_2E}{\sqrt{15}}\left(\begin{array}{cc}
3 & -\sqrt{\frac{5}{2}}\\
-\sqrt{\frac{5}{2}} & 1
\end{array}\right)\nonumber
\end{eqnarray}
with the same eigenvalues as the previous one:
\begin{eqnarray}
M_{1,2}^{\left(3,2,\frac{1}{3}\right)} = \frac{E\lambda_2}{2\sqrt{15}}\left[4\pm \sqrt{14}\right] \neq 0
\end{eqnarray}
\item (1,3,0) : There are also two of them and the mass matrix is
\begin{eqnarray}
&& {\rm c}~:~\widehat A_{(1,3,1)}^{(1,3,0)},~\widehat E_{(1,3,3)}^{(1,3,0)};~~
{\rm r}~:~\widehat A_{(1,3,1)}^{(1,3,0)},~\widehat E_{(1,3,3)}^{(1,3,0)}\nonumber\\
&&\left(\begin{array}{cc}
m_1+\frac{3\lambda_2E}{\sqrt{15}} & \lambda_2A_1\\
\lambda_2A_1 & m_2+\frac{9\lambda_1E}{\sqrt{15}}
\end{array}\right)=\frac{E}{\sqrt{15}}\left(\begin{array}{cc}
5\lambda_2 & 0\\
0 & \lambda_2+\frac{15}{2}\lambda_1
\end{array}\right)\nonumber
\end{eqnarray}
So the mass eigenvalues are
\begin{eqnarray}
 M_1^{(1,3,0)}=\frac{5E}{\sqrt{15}}\lambda_2\neq
0,~~M_2^{(1,3,0)}=\frac{E}{\sqrt{15}}\left(\lambda_2+\frac{15}{2}\lambda_1\right)\neq 0
\end{eqnarray}
\item $[(1,3,2)+ {\rm c.c.}]$ : There is only one such multiplet whose eigenvalue is given by
\begin{eqnarray}
&&{\rm c}~:~\widehat E_{(1,3,3)}^{(1,3,2)},~~{\rm r}~:~\widehat E_{(1,3,3)}^{(1,3,-2)}\nonumber\\
&&M^{(1,3,2)} = m_2+\frac{9}{\sqrt{15}}\lambda_1E = \frac{E}{\sqrt{15}}\left(\lambda_2+\frac{15}{2}\lambda_1\right)\neq 0
\end{eqnarray}
\item $\left[\left(6,1,-\frac{4}{3}\right)+{\rm c.c.}\right]$ : Its eigenvalue is
\begin{eqnarray}
 &&{\rm c}~:~\widehat E_{(20',1,1)}^{\left(6,1,-\frac{4}{3}\right)},~~{\rm r}~:~\widehat
E_{(20',1,1)}^{\left(6,1,\frac{4}{3}\right)}\nonumber\\
 &&M^{\left(6,1,-\frac{4}{3}\right)} = m_2-\frac{6}{\sqrt{15}}\lambda_1E =
\frac{E}{\sqrt{15}}\left(\lambda_2-\frac{15}{2}\lambda_1\right)\neq 0
\end{eqnarray}
 unless $\lambda_2=\frac{15}{2}\lambda_1$ (which we assume not to be the case).
\item (8,1,0) : There are two of them and the mass matrix is
\begin{eqnarray}
&&{\rm c} : \widehat A_{(15,1,1)}^{(8,1,0)},~~\widehat E_{(20',1,1)}^{(8,1,0)};~~
{\rm r} : \widehat A_{(15,1,1)}^{(8,1,0)},~~\widehat E_{(20',1,1)}^{(8,1,0)}\nonumber\\
&&\left(\begin{array}{cc}
m_1-\frac{2\lambda_2E}{\sqrt{15}} & \sqrt{\frac{2}{3}}\lambda_2A_2\\
\sqrt{\frac{2}{3}}\lambda_2A_2 & m_2-\frac{6\lambda_1E}{\sqrt{15}}
\end{array}\right)=\frac{E}{\sqrt{15}}\left(\begin{array}{cc}
0 & \sqrt{\frac{45}{2}}\lambda_2\\
\sqrt{\frac{45}{2}}\lambda_2 & \lambda_2-\frac{15}{2}\lambda_1
\end{array}\right)\nonumber
\end{eqnarray}
with the mass eigenvalues
\begin{eqnarray}
 M_{1,2}^{(8,1,0)}=\frac{E}{2\sqrt{15}}\left[\left(\lambda_2-\frac{15}{2}\lambda_1\right)^2\pm
\sqrt{\left(\lambda_2-\frac{15}{2}\lambda_1\right)^2+90\lambda_2^2}~\right]\neq 0
\end{eqnarray}
\end{itemize}
 Thus we see that all the other multiplets have non-zero masses, and moreover,
all these masses are of order $E\sim M_G$. Hence, none of these multiplets
 will contribute to the running of gauge coupling up to the unification scale
$M_G$ except the color triplets since
these color triplets have masses of order of the SUSY breaking scale.

Note that the $\mathbf {10}_H$-Higgs field also has a color triplet pair
$\left[\left(3,1,-\frac23\right)+{\rm c.c.}\right]$ under the SM gauge group,
apart from the TeV-scale bi-doublet fields $\Phi_{1,2}$ used in the SUSYLR
model in Section 4 which
reduce to $(1,2,\pm 1)$ under the SM gauge group. At the GUT-scale,
the $\mathbf H\equiv \mathbf {10}_H$ field interacts with the
$\mathbf E\equiv \mathbf {54}_H$ field
by the following term in the superpotential:
\begin{eqnarray}
W_{10}=\frac12 m_3 \mathbf H^2+\lambda_3 \mathbf E \mathbf H^2
\end{eqnarray}
After the $\mathbf {54}_H$ acquires a VEV, this gives rise to the color
triplet mass
\begin{eqnarray}
&&{\rm c}~:~\widehat{H}_{(6,1,1)}^{\left(3,1,-\frac23\right)};~~
{\rm r}~:~\widehat H_{(6,1,1)}^{\left(\bar{3},1,\frac23\right)}\nonumber\\
&&M^{\left(3,1,\frac23\right)} = m_3-\frac{2\lambda_3E}{\sqrt{15}}
\end{eqnarray}
while the doublet mass is
\begin{eqnarray}
&&{\rm c}~:~\widehat H_{(1,2,2)}^{(1,2,1)};~~{\rm r}~:~\widehat H_{(1,2,2)}^{(1,2,-1)}\nonumber\\
&&M^{(1,2,1)} = m_3+\sqrt{\frac35}\lambda_3E
\end{eqnarray}
We see that the $(1,2,\pm 1)$ field can be made light by fine-tuning
$m_3+\sqrt{\frac35}\lambda_3E \sim$ TeV which still leaves the
$\left(3,1,\frac23\right)$ field heavy (of order $M_G$).

Finally, let us discuss how only the right handed doublets fields $(\chi^c,~
\overline{\chi}^c)$ from ${\mathbf 16}_H$-Higgs fields
($\psi_H$) remain massless at the GUT scale. Note that in the left-right
language, the fields in {\bf 16} are
$Q_H~(3,2,1,\frac13)\oplus Q^c_H~(\bar{3},1,2,-\frac13)$ and
$\chi~(1,2,1,-1)\oplus\chi^c(1,1,2,+1)$, and similarly for
$\overline{\bf 16}_H\equiv \bar{\psi}_H$ field. The superpotential involving
these fields is
\begin{equation}
W_{16}~=~M_{16}\bar{\psi}_H\psi_H+\lambda \bar{\psi}_H A \psi_H
\end{equation}
The second coupling has been worked out explicitly in Ref.~\cite{aulakh}.
On substituting the VEV of the {\bf 45}-Higgs field ($A$), we get the
following masses for the $Q_H~(3,2,1,\frac13)\oplus
Q^c_H~(\bar{3},1,2,-\frac13)$ and
$\chi~(1,2,1,-1)\oplus\chi^c(1,1,2,+1)$ fields:
\begin{eqnarray}
&&M_{Q_H-\overline{Q}_H}=M_{16}+\lambda A_2; ~~
M_{Q^c_H-\overline{Q}^c_H}=M_{16}-\lambda A_2;\nonumber\\
&&M_{\chi-\overline{\chi}}=M_{16}-3\lambda A_2; ~~
M_{\chi^c-\overline{\chi}^c}=M_{16}+3\lambda A_2
\end{eqnarray}
From this we see that to get only the $\chi^c$ fields light, we have to
fine-tune $M_{16}+3\lambda A_2\sim $ TeV.
With this assumption, all other fields remain heavy at the GUT scale.

\section*{Appendix B: RGEs for fermion masses and mixing}
Given the form of the bi-doublets VEVs as in Eq.~(\ref{eq:bivev}), it immediately follows from the first two terms of the
superpotential Eq.~(\ref{eq:modsup}) that the fermion mass matrices can be written as
\begin{eqnarray}
M_u=\frac{1}{\sqrt 2}v_uh_2,~~M_d=\frac{1}{\sqrt 2}v_dh_1,~~M_e=\frac{1}{\sqrt 2}v_dh'_1,~~{\rm and}~M_D=\frac{1}{\sqrt 2}v_uh'_2
\end{eqnarray}
Henceforth, for clarity, we will denote the Yukawa couplings as
\[h_U\equiv h_2,~~h_D\equiv h_1,~~h_E\equiv h'_1,~~h_N\equiv h'_2\]
Then using Eqs.~(\ref{eq:ha}, \ref{eq:hap}) and (\ref{eq:vu}, \ref{eq:vd})
the RGEs for the fermion mass matrices can be written as
\begin{eqnarray}
16\pi^2\frac{dM_u}{dt} &=& M_u\left[4h_U^\dagger h_U+2h_D^\dagger h_D-\sum_i \tilde{C}_i^{(q)}g_i^2\right]\nonumber\\
&&+M_d\tan\beta\left[{\rm Tr}\left(3h_D^\dagger h_U+h_E^\dagger h_N\right)+2h_D^\dagger h_U+C^\Phi_{12}\right] \label{eq:mu1}\\
16\pi^2\frac{dM_d}{dt} &=& M_d\left[4h_D^\dagger h_D+2h_U^\dagger h_U-\sum_i \tilde{C}_i^{(q)}g_i^2\right]\nonumber\\
&&+\frac{M_u}{\tan\beta}\left[{\rm Tr}\left(3h_U^\dagger h_D+h_N^\dagger h_E\right)+2h_U^\dagger h_D+C^\Phi_{21}\right]\label{eq:md1}\\
16\pi^2\frac{dM_e}{dt} &=& M_e\left[4h_E^\dagger h_E+2h_N^\dagger h_N+C^\chi-\sum_i \tilde{C}_i^{(l)}g_i^2\right]\nonumber\\
&&+\frac{M_D}{\tan\beta}\left[{\rm Tr}\left(3h_U^\dagger h_D+h_N^\dagger h_E\right)+2h_N^\dagger h_E+C^\Phi_{21}\right]\label{eq:me1}\\
16\pi^2\frac{dM_D}{dt} &=& M_D\left[4h_N^\dagger h_N+2h_E^\dagger h_E+C^\chi-\sum_i \tilde{C}_i^{(l)}g_i^2\right]\nonumber\\
&&+M_e\tan\beta\left[{\rm Tr}\left(3h_D^\dagger h_U+h_E^\dagger h_N\right)+2h_E^\dagger h_N+C^\Phi_{12}\right]\label{eq:mnu1}
\end{eqnarray}
where $C^\Phi_{ab}=4\left(\mu^{\Phi^\dagger}_\alpha\mu^\Phi_\alpha\right)_{ab},~~
C^\chi=\left(\mu^{L^c}\right)^\dagger_{\alpha q}\mu^{L^c}_{\alpha q}$, and for $i=\mathbf 3_c,~\mathbf 2_L,~\mathbf 2_R,~\mathbf 1_{B-L}$,
\begin{eqnarray}
\tilde{C}_i^{(q)} = \left(\frac{16}{3},\frac{3}{2},\frac{3}{2},\frac{1}{6}\right),~~\tilde{C}_i^{(l)} = \left(0,\frac{3}{2},\frac{3}{2},\frac{3}{2}\right)
\end{eqnarray}
Note that the second line in each of the above mass RGEs, Eqs.~
(\ref{eq:mu1}-\ref{eq:mnu1}), is characteristic of the left-right models, and
does not appear in MSSM.

Not all the parameters of the Yukawa matrices are physical. Under an arbitrary unitary transformation on
the left(right)-handed fermion fields, ${\cal F}_{L(R)}\to L(R)_f{\cal F}_{L(R)}$ (where ${\cal F}=U,D,E,N$), the Yukawa matrices undergo a bi-unitary transformation,
$h_f\to L_f h_f R_f^\dagger$ and the charged current becomes off-diagonal, with the CKM
mixing matrix $L_U L_D^\dagger$. We will also have a leptonic counterpart of the CKM matrix that represents the mixing between the charged lepton and Dirac neutrino sector. However, as the running of lepton masses is very mild and we are working only to the one-loop order, we can safely ignore this mixing in the leptonic sector. Moreover, if we assume the $CP$ phase in the Higgs VEV to be zero, then the mass
matrices are Hermitian and $L_f=R_f$ (manifest left-right). Thus we may perform scale-dependent unitary transformations
$L_f(\mu)$ on the fermion bases so as to diagonalize the
Yukawa matrices, and hence the mass matrices, at each scale:
\begin{equation}
\widehat{h}_f(\mu)=L_f(\mu)h_f(\mu)L_f^\dagger(\mu),~{\rm and}~\widehat{M}_f=L_f(\mu)M_f(\mu)L_f^\dagger(\mu),
\label{eq:hat}
\end{equation}
where $\widehat{h}_f$ and $\widehat{M}_f$ denote the diagonalized Yukawa and mass matrices, respectively.

The RGEs for the physically relevant quantities, namely the mass eigenvalues
$\widehat{M}_f(\mu)$ and the scale-dependent CKM matrix $V_{\rm CKM}(\mu)=L_U(\mu)L_D^\dagger(\mu)$, are both
contained in the RGEs of $\widehat{M}_f^2(\mu)=L_f^\dagger(\mu)M_f(\mu)M_f^\dagger(\mu)L_f(\mu)$:
\begin{eqnarray}
\frac{d}{dt}\left(\widehat{M}_u^2\right) &=& \left[\dot{L}_UL_U^\dagger,\widehat{M}_u^2\right]+\frac{1}{16\pi^2}\left[4\widehat{h}_U^2+2\widehat{h}_D^2-\sum_i \tilde{C}_i^{(q)}g_i^2\right]2\widehat{M}_u^2\nonumber\\
&&+\frac{1}{16\pi^2}\tan\beta\left[\left\{{\rm Tr}\left(3V_{\rm CKM}\widehat{h}_DV_{\rm CKM}^\dagger \widehat{h}_U\right)
+C^\Phi_{12}\right\}\left(V_{\rm CKM}\widehat{M}_dV_{\rm CKM}^\dagger \widehat{M}_u\right)\right.\nonumber\\
&&\left.+2V_{\rm CKM}\widehat{M}_d\widehat{h}_DV_{\rm CKM}^\dagger \widehat{h}_U\widehat{M}_u+{\rm h.c.}\right]\label{eq:mu2}\\
\frac{d}{dt}\left(\widehat{M}_d^2\right) &=& \left[\dot{L}_DL_D^\dagger,\widehat{M}_d^2\right]+\frac{1}{16\pi^2}\left[4\widehat{h}_D^2+2\widehat{h}_U^2-\sum_i \tilde{C}_i^{(q)}g_i^2\right]2\widehat{M}_d^2\nonumber\\
&&+\frac{1}{16\pi^2}\frac{1}{\tan\beta}\left[\left\{{\rm Tr}\left(3V_{\rm CKM}\widehat{h}_DV_{\rm CKM}^\dagger \widehat{h}_U\right)
+C^\Phi_{12}\right\}\left(\widehat{M}_dV_{\rm CKM}^\dagger\widehat{M}_uV_{\rm CKM}\right)\right.\nonumber\\
&&\left.+2\widehat{M}_d\widehat{h}_DV_{\rm CKM}^\dagger\widehat{h}_U\widehat{M}_uV_{\rm CKM}+{\rm h.c.}\right]\label{eq:md2}\\
\frac{d}{dt}\left(\widehat{M}_e^2\right) &=& \left[\dot{L}_EL_E^\dagger,\widehat{M}_e^2\right]+\frac{1}{16\pi^2}\left[4\widehat{h}_E^2+2\widehat{h}_N^2+\Re\left(C^\chi\right)-\sum_i \tilde{C}_i^{(l)}g_i^2\right]2\widehat{M}_e^2\label{eq:me2}\\
\frac{d}{dt}\left(\widehat{M}_D^2\right) &=& \left[\dot{L}_NL_N^\dagger,\widehat{M}_D^2\right]+\frac{1}{16\pi^2}\left[4\widehat{h}_N^2+2\widehat{h}_E^2+\Re\left(C^\chi\right)-\sum_i \tilde{C}_i^{(l)}g_i^2\right]2\widehat{M}_D^2\label{eq:mnu2}
\end{eqnarray}
where $\dot{L}\equiv \frac{dL}{dt}$ and $\Re\left(C^\chi\right)$ denotes the real part of $C^\chi$.
The commutator $\left[\dot{L}_f L_f^\dagger,\widehat{M}_f^2\right]$ has vanishing diagonal elements
because $\widehat{M}_f^2$ is diagonal. Thus the RGEs for the mass eigenvalues $m_f^2$ follow immediately
from the diagonal entries of Eqs.~(\ref{eq:mu2}-\ref{eq:mnu2}). Using dominance of Yukawa couplings of the
third generation over the first two, i.e.
\[h_t^2\gg h_c^2 \gg h_u^2,~~h_b^2\gg h_s^2 \gg h_d^2,~~h_\tau^2 \gg h_\mu^2 \gg h_e^2,~~h_{N_3}^2 \gg h_{N_2}^2 \gg h_{N_1}^2,\]
we obtain the following RGEs for the mass eigenvalues of the fermions:
\begin{eqnarray}
16\pi^2\frac{dm_u}{dt} &\simeq & \left(4h_u^2+2h_d^2-\sum_i \tilde{C}_i^{(q)}g_i^2\right)m_u + \tan\beta\left[3|V_{tb}|^2h_bh_t+r_q\right]\sum_{j=d,s,b}|V_{uj}|^2m_j\nonumber \\
16\pi^2\frac{dm_c}{dt} &\simeq & \left(4h_c^2+2h_s^2-\sum_i \tilde{C}_i^{(q)}g_i^2\right)m_c + \tan\beta\left[3|V_{tb}|^2h_bh_t+r_q\right]\sum_{j=d,s,b}|V_{cj}|^2m_j\nonumber\\
16\pi^2\frac{dm_t}{dt} &\simeq & \left(4h_t^2+2h_b^2-\sum_i \tilde{C}_i^{(q)}g_i^2\right)m_t + \tan\beta\left[\left(3|V_{tb}|^2+2\right)h_bh_t+r_q\right]|V_{tb}|^2m_b\nonumber\\
16\pi^2\frac{dm_d}{dt} &\simeq & \left(4h_d^2+2h_u^2-\sum_i \tilde{C}_i^{(q)}g_i^2\right)m_d + \frac{1}{\tan\beta}\left[3|V_{tb}|^2h_bh_t+r_q\right]\sum_{j=u,c,t}|V_{jd}|^2m_j\nonumber\\
16\pi^2\frac{dm_s}{dt} &\simeq & \left(4h_s^2+2h_c^2-\sum_i \tilde{C}_i^{(q)}g_i^2\right)m_s + \frac{1}{\tan\beta}\left[3|V_{tb}|^2h_bh_t+r_q\right]\sum_{j=u,c,t}|V_{js}|^2m_j\nonumber\\
16\pi^2\frac{dm_b}{dt} &\simeq & \left(4h_b^2+2h_t^2-\sum_i \tilde{C}_i^{(q)}g_i^2\right)m_b + \frac{1}{\tan\beta}\left[\left(3|V_{tb}|^2+2\right)h_bh_t+r_q\right]|V_{tb}|^2m_t\nonumber\\
16\pi^2\frac{dm_e}{dt} &\simeq & \left(4h_e^2+2h_{N_1}^2+r_l-\sum_i \tilde{C}_i^{(l)}g_i^2\right)m_e\nonumber\\
16\pi^2\frac{dm_\mu}{dt} &\simeq & \left(4h_\mu^2+2h_{N_2}^2+r_l-\sum_i \tilde{C}_i^{(l)}g_i^2\right)m_\mu\nonumber\\
16\pi^2\frac{dm_\tau}{dt} &\simeq & \left(4h_\tau^2+2h_{N_3}^2+r_l-\sum_i \tilde{C}_i^{(l)}g_i^2\right)m_\tau\nonumber\\
16\pi^2\frac{dm_{N_1}}{dt} &\simeq & \left(4h_{N_1}^2+2h_e^2+r_l-\sum_i \tilde{C}_i^{(l)}g_i^2\right)m_{N_1}\nonumber\\
16\pi^2\frac{dm_{N_2}}{dt} &\simeq & \left(4h_{N_2}^2+2h_\mu^2+r_l-\sum_i \tilde{C}_i^{(l)}g_i^2\right)m_{N_2}\nonumber\\
16\pi^2\frac{dm_{N_3}}{dt} &\simeq & \left(4h_{N_3}^2+2h_\tau^2+r_l-\sum_i \tilde{C}_i^{(l)}g_i^2\right)m_{N_3}
\label{eq:rgmass}
\end{eqnarray}
where $r_q=\Re\left(C^\Phi_{12}\right)$ and $r_l=\Re\left(C^\chi\right)$.

The VEV RGEs, Eqs. (\ref{eq:vd}) and (\ref{eq:vu}), for third generation dominance become
\begin{eqnarray}
	16\pi^2\frac{dv_u}{dt}&\simeq & v_u\left[\frac{3}{2}g_{2L}^2+
	\frac{3}{2}g_{2R}^2-3h_t^2-h_{N_3}^2 - C^\Phi_{22}\right],\\
	16\pi^2\frac{dv_d}{dt}&\simeq & v_d\left[\frac{3}{2}g_{2L}^2+
	\frac{3}{2}g_{2R}^2-3h_b^2-h_\tau^2 - C^\Phi_{11}\right]
\end{eqnarray}

The RGE for the CKM matrix $V_{\rm CKM}=L_UL_D^\dagger$ is given by
\begin{eqnarray}
\frac{d}{dt}V_{\rm CKM} &=& \dot{L}_UL_D^\dagger + L_U\dot{L}_D^\dagger
= \dot{L}_UL_U^\dagger V_{\rm CKM} - V_{\rm CKM} \dot{L}_D L_D^\dagger,\nonumber\\
{\rm or},~~\frac{d}{dt}V_{\alpha\beta} &=& \sum_{\gamma=u,c,t}\left(\dot{L}_UL_U^\dagger\right)_{\alpha\gamma}V_{\gamma\beta}
- \sum_{\gamma=d,s,b}V_{\alpha\gamma}\left(\dot{L}_DL_D^\dagger\right)_{\gamma\beta}
\label{eq:ckm1}
\end{eqnarray}
However, the diagonal elements of $\dot{L}_{U,D}L_{U,D}^\dagger$ are not determined by Eqs.~(\ref{eq:mu2}) and (\ref{eq:md2}). This is because
Eq.~(\ref{eq:hat}) determines $L_{U,D}$ only up to right multiplication by a diagonal matrix of scale-dependent phases.
These undetermined phases contribute arbitrary imaginary functions to the diagonal elements of $\dot{L}_{U,D}L_{U,D}^\dagger$.
But the off-diagonal elements are unambiguously determined because they receive no contribution from the phases. We can, nevertheless,
make the diagonal entries of $\dot{L}_{U,D}L_{U,D}^\dagger$, which are manifestly imaginary, vanish by an appropriate choice of phases.
With this choice of phases, we can then obtain the RGEs for the CKM matrix elements using Eq.~(\ref{eq:ckm1}):
\begin{eqnarray}
\frac{d}{dt}V_{\alpha\beta} &=&  \sum_{\stackrel{\scriptstyle \gamma=u,c,t}{\gamma\neq\alpha}}\left(\dot{L}_UL_U^\dagger\right)_{\alpha\gamma}V_{\gamma\beta}
- \sum_{\stackrel{\scriptstyle \gamma=d,s,b}{\gamma\neq\beta}}V_{\alpha\gamma}\left(\dot{L}_DL_D^\dagger\right)_{\gamma\beta}\nonumber\\
&=& \frac{1}{16\pi^2}\left(\sum_{\stackrel{\scriptstyle \gamma=u,c,t}{\gamma\neq\alpha}}\left[\frac{\tan\beta}{m_\alpha-m_\gamma}\left\{{\rm Tr}\left(3V\widehat{h}_DV^\dagger \widehat{h}_U\right)+r_q\right\}\left(V\widehat{M}_d V^\dagger\right)_{\alpha\gamma}\right.\right.\nonumber\\
&&\left.\left.~~~~~~~~~~+ \frac{4}{v_d^2}\frac{m_\alpha^2+m_\gamma^2}{m_\alpha^2-m_\gamma^2}\left(V\widehat{M}_d^2V^\dagger\right)_{\alpha\gamma}\right]V_{\gamma\beta} \right. \nonumber\\ && \left.
- \sum_{\stackrel{\scriptstyle \gamma=d,s,b}{\gamma\neq\beta}}V_{\alpha\gamma}\left[\frac{1}{\tan\beta(m_\gamma-m_\beta)}
\left\{{\rm Tr}\left(3V\widehat{h}_DV^\dagger \widehat{h}_U\right)+r_q\right\}\left(V^\dagger \widehat{M}_u V\right)_{\gamma\beta} \right.\right.\nonumber\\&&\left.\left.
~~~~~~~~~~+ \frac{4}{v_u^2}\frac{m_\gamma^2+m_\beta^2}{m_\gamma^2-m_\beta^2}\left(V^\dagger \widehat{M}_u^2V\right)_{\gamma\beta}\right]\right)
\end{eqnarray}
As before, we use the third generation dominance and get the following RGEs for $V_{\alpha\beta}$:
\begin{eqnarray}
16\pi^2\frac{d}{dt}V_{ud} &\simeq &
-\tan\beta\left(3|V_{tb}|^2h_bh_t+r_q\right)\left[\frac{\left(V\widehat{M}_dV^\dagger\right)_{uc}V_{cd}}{m_c}
+\frac{\left(V\widehat{M}_dV^\dagger\right)_{ut}V_{td}}{m_t}\right]\nonumber\\
&& -\frac{4}{v_d^2}\left[\left(V\widehat{M}_d^2V^\dagger\right)_{uc}V_{cd}
+\left(V\widehat{M}_d^2V^\dagger\right)_{ut}V_{td}\right]\nonumber\\
&& -\frac{1}{\tan\beta}\left(3|V_{tb}|^2h_bh_t+r_q\right)
\left[\frac{V_{us}\left(V^\dagger\widehat{M}_uV\right)_{sd}}{m_s}
+\frac{V_{ub}\left(V^\dagger\widehat{M}_uV\right)_{bd}}{m_b}\right]\nonumber\\
&& -\frac{4}{v_u^2}\left[V_{us}\left(V^\dagger\widehat{M}_u^2V\right)_{sd}
+V_{ub}\left(V^\dagger \widehat{M}_u^2V\right)_{bd}\right]\nonumber \\
16\pi^2\frac{d}{dt}V_{us} &\simeq &
-\tan\beta\left(3|V_{tb}|^2h_bh_t+r_q\right)\left[\frac{\left(V\widehat{M}_dV^\dagger\right)_{uc}V_{cs}}{m_c}
+\frac{\left(V\widehat{M}_dV^\dagger\right)_{ut}V_{ts}}{m_t}\right]\nonumber\\
&& -\frac{4}{v_d^2}\left[\left(V\widehat{M}_d^2V^\dagger\right)_{uc}V_{cs}
+\left(V\widehat{M}_d^2V^\dagger\right)_{ut}V_{ts}\right]\nonumber\\
&& -\frac{1}{\tan\beta}\left(3|V_{tb}|^2h_bh_t+r_q\right)
\left[-\frac{V_{ud}\left(V^\dagger\widehat{M}_uV\right)_{ds}}{m_s}
+\frac{V_{ub}\left(V^\dagger\widehat{M}_uV\right)_{bs}}{m_b}\right]\nonumber\\
&& -\frac{4}{v_u^2}\left[-V_{ud}\left(V^\dagger\widehat{M}_u^2V\right)_{ds}
+V_{ub}\left(V^\dagger \widehat{M}_u^2V\right)_{bs}\right]\nonumber \\
16\pi^2\frac{d}{dt}V_{ub} &\simeq &
-\tan\beta\left(3|V_{tb}|^2h_bh_t+r_q\right)\left[\frac{\left(V\widehat{M}_dV^\dagger\right)_{uc}V_{cb}}{m_c}
+\frac{\left(V\widehat{M}_dV^\dagger\right)_{ut}V_{tb}}{m_t}\right]\nonumber\\
&& -\frac{4}{v_d^2}\left[\left(V\widehat{M}_d^2V^\dagger\right)_{uc}V_{cb}
+\left(V\widehat{M}_d^2V^\dagger\right)_{ut}V_{tb}\right]\nonumber\\
&& +\frac{1}{m_b\tan\beta}\left(3|V_{tb}|^2h_bh_t+r_q\right)
\left[V_{ud}\left(V^\dagger\widehat{M}_uV\right)_{db}
+V_{us}\left(V^\dagger\widehat{M}_uV\right)_{sb}\right]\nonumber\\
&& +\frac{4}{v_u^2}\left[V_{ud}\left(V^\dagger\widehat{M}_u^2V\right)_{db}
+V_{us}\left(V^\dagger \widehat{M}_u^2V\right)_{sb}\right]\nonumber \\
16\pi^2\frac{d}{dt}V_{cd} &\simeq &
-\tan\beta\left(3|V_{tb}|^2h_bh_t+r_q\right)\left[
-\frac{\left(V\widehat{M}_dV^\dagger\right)_{cu}V_{ud}}{m_c}
+\frac{\left(V\widehat{M}_dV^\dagger\right)_{ct}V_{td}}{m_t}\right]\nonumber\\
&& -\frac{4}{v_d^2}\left[-\left(V\widehat{M}_d^2V^\dagger\right)_{cu}V_{ud}
+\left(V\widehat{M}_d^2V^\dagger\right)_{ct}V_{td}\right]\nonumber\\
&& -\frac{1}{\tan\beta}\left(3|V_{tb}|^2h_bh_t+r_q\right)\left[
\frac{V_{cs}\left(V^\dagger\widehat{M}_uV\right)_{sd}}{m_s}
+\frac{V_{cb}\left(V^\dagger\widehat{M}_uV\right)_{bd}}{m_b}\right]\nonumber\\
&& -\frac{4}{v_u^2}\left[V_{cs}\left(V^\dagger\widehat{M}_u^2V\right)_{sd}
+V_{cb}\left(V^\dagger \widehat{M}_u^2V\right)_{bd}\right]\nonumber \\
16\pi^2\frac{d}{dt}V_{cs} &\simeq &
-\tan\beta\left(3|V_{tb}|^2h_bh_t+r_q\right)\left[
-\frac{\left(V\widehat{M}_dV^\dagger\right)_{cu}V_{us}}{m_c}
+\frac{\left(V\widehat{M}_dV^\dagger\right)_{ct}V_{ts}}{m_t}\right]\nonumber\\
&& -\frac{4}{v_d^2}\left[-\left(V\widehat{M}_d^2V^\dagger\right)_{cu}V_{us}
+\left(V\widehat{M}_d^2V^\dagger\right)_{ct}V_{ts}\right]\nonumber\\
&& -\frac{1}{\tan\beta}\left(3|V_{tb}|^2h_bh_t+r_q\right)\left[
-\frac{V_{cd}\left(V^\dagger\widehat{M}_uV\right)_{ds}}{m_s}
+\frac{V_{cb}\left(V^\dagger\widehat{M}_uV\right)_{bs}}{m_b}\right]\nonumber\\
&& -\frac{4}{v_u^2}\left[
-V_{cd}\left(V^\dagger\widehat{M}_u^2V\right)_{ds}
+V_{cb}\left(V^\dagger \widehat{M}_u^2V\right)_{bs}\right]\nonumber \\
16\pi^2\frac{d}{dt}V_{cb} &\simeq &
-\tan\beta\left(3|V_{tb}|^2h_bh_t+r_q\right)\left[
-\frac{\left(V\widehat{M}_dV^\dagger\right)_{cu}V_{ub}}{m_c}
+\frac{\left(V\widehat{M}_dV^\dagger\right)_{ct}V_{tb}}{m_t}\right]\nonumber\\
&& -\frac{4}{v_d^2}\left[-\left(V\widehat{M}_d^2V^\dagger\right)_{cu}V_{ub}
+\left(V\widehat{M}_d^2V^\dagger\right)_{ct}V_{tb}\right]\nonumber\\
&& +\frac{1}{m_b\tan\beta}\left(3|V_{tb}|^2h_bh_t+r_q\right)\left[
V_{cd}\left(V^\dagger\widehat{M}_uV\right)_{db}
+V_{cs}\left(V^\dagger\widehat{M}_uV\right)_{sb}\right]\nonumber\\
&& +\frac{4}{v_u^2}\left[
V_{cd}\left(V^\dagger\widehat{M}_u^2V\right)_{db}
+V_{cs}\left(V^\dagger \widehat{M}_u^2V\right)_{sb}\right]\nonumber \\
16\pi^2\frac{d}{dt}V_{td} &\simeq &
\frac{\tan\beta}{m_t}\left(3|V_{tb}|^2h_bh_t+r_q\right)\left[
\left(V\widehat{M}_dV^\dagger\right)_{tu}V_{ud}
+\left(V\widehat{M}_dV^\dagger\right)_{tc}V_{cd}\right]\nonumber\\
&& +\frac{4}{v_d^2}\left[\left(V\widehat{M}_d^2V^\dagger\right)_{tu}V_{ud}
+\left(V\widehat{M}_d^2V^\dagger\right)_{tc}V_{cd}\right]\nonumber\\
&& -\frac{1}{\tan\beta}\left(3|V_{tb}|^2h_bh_t+r_q\right)\left[
\frac{V_{ts}\left(V^\dagger\widehat{M}_uV\right)_{sd}}{m_s}
+\frac{V_{tb}\left(V^\dagger\widehat{M}_uV\right)_{bd}}{m_b}\right]\nonumber\\
&& -\frac{4}{v_u^2}\left[V_{ts}\left(V^\dagger\widehat{M}_u^2V\right)_{sd}
+V_{tb}\left(V^\dagger \widehat{M}_u^2V\right)_{bd}\right]\nonumber \\
16\pi^2\frac{d}{dt}V_{ts} &\simeq &
\frac{\tan\beta}{m_t}\left(3|V_{tb}|^2h_bh_t+r_q\right)\left[
\left(V\widehat{M}_dV^\dagger\right)_{tu}V_{us}
+\left(V\widehat{M}_dV^\dagger\right)_{tc}V_{cs}\right]\nonumber\\
&& +\frac{4}{v_d^2}\left[\left(V\widehat{M}_d^2V^\dagger\right)_{tu}V_{us}
+\left(V\widehat{M}_d^2V^\dagger\right)_{tc}V_{cs}\right]\nonumber\\
&& -\frac{1}{\tan\beta}\left(3|V_{tb}|^2h_bh_t+r_q\right)\left[
-\frac{V_{td}\left(V^\dagger\widehat{M}_uV\right)_{ds}}{m_s}
+\frac{V_{tb}\left(V^\dagger\widehat{M}_uV\right)_{bs}}{m_b}\right]\nonumber\\
&& -\frac{4}{v_u^2}\left[-V_{td}\left(V^\dagger\widehat{M}_u^2V\right)_{ds}
+V_{tb}\left(V^\dagger \widehat{M}_u^2V\right)_{bs}\right]\nonumber\\
16\pi^2\frac{d}{dt}V_{tb} &\simeq &
\frac{\tan\beta}{m_t}\left(3|V_{tb}|^2h_bh_t+r_q\right)\left[
\left(V\widehat{M}_dV^\dagger\right)_{tu}V_{ub}
+\left(V\widehat{M}_dV^\dagger\right)_{tc}V_{cb}\right]\nonumber\\
&& +\frac{4}{v_d^2}\left[\left(V\widehat{M}_d^2V^\dagger\right)_{tu}V_{ub}
+\left(V\widehat{M}_d^2V^\dagger\right)_{tc}V_{cb}\right]\nonumber\\
&& +\frac{1}{m_b\tan\beta}\left(3|V_{tb}|^2h_bh_t+r_q\right)\left[
V_{td}\left(V^\dagger\widehat{M}_uV\right)_{db}
+V_{ts}\left(V^\dagger\widehat{M}_uV\right)_{sb}\right]\nonumber\\
&& +\frac{4}{v_u^2}\left[
V_{td}\left(V^\dagger\widehat{M}_u^2V\right)_{db}
+V_{ts}\left(V^\dagger \widehat{M}_u^2V\right)_{sb}\right]
\label{eq:rgckm}
\end{eqnarray}
We have presented the results for these RGEs even though they look quite messy 
because we believe this is the first time such an analysis has been carried 
out in the SUSYLR model, and these analytical results at the one-loop level 
may be useful later for future work in this direction.

In order to solve these mass and mixing RGEs numerically, we need to know the initial values for all the 23 variables
(12 masses, 9 CKM elements and 2 VEVs). We know the experimental values at
$\tilde{\mu}=M_Z$ for all of them except for the
Dirac neutrino masses $m_{N_i}$. We fix these values by iterations using the GUT-scale predicted values, $m_{N_i}(M_G)$,
which, in turn, are determined completely in terms of the other fermion masses at the GUT-scale in $SO(10)$ GUT models. Here
we note that adjusting the GUT-scale values of $m_{N_i}$ to fit the $SO(10)$ model prediction do not change the other fermion masses at this scale
significantly even though they are all coupled equations because of the mild running of the neutrino masses. Hence the
mass and mixing values given in Eqs.~(\ref{eq:GUTmass}) can be considered as generic and independent of the specific $SO(10)$ model chosen.

We also have the free parameters $r_q$ and $r_l$ corresponding to the couplings $\mu_{\alpha}^{\Phi}$ and $\mu_{\alpha q}^{L^c}$.
Assuming the couplings $\mu_\alpha$ to be the same $\forall~\alpha=1,2,3$, we have
\begin{eqnarray}
C^\Phi_{ab} &=& 4\left(\mu^{\Phi^\dagger}_\alpha \mu^\Phi_\alpha\right)_{ab} = 12\left(\mu^{\Phi^\dagger}\mu^\Phi\right)_{ab}
= 12\sum_{c=1}^2 \mu^{\Phi^*}_{ca}\mu^\Phi_{cb}\nonumber\\
C^\chi &=& \left(\mu^{L^c}\right)^\dagger_{\alpha q}\mu^{L^c}_{\alpha q}
= 3\left[\left(\mu^{L^c}\right)^*_q\mu^{L^c}_q\right]\nonumber
\end{eqnarray}
Further assuming $\mu^\Phi_{ab}=\mu_\phi ~\forall~ a,~b=1,2$ and $\mu^{L^c}_q=\mu_l ~\forall~ q=1,2$, we have
\begin{eqnarray}
r_q= 24|\mu_\phi|^2,~r_l=6|\mu_l|^2\nonumber
\end{eqnarray}
where $\mu_\phi$ and $\mu_l$ can take values between 0 and 1 (for the theory to remain perturbative).

For the running behavior shown in Figures 2,
we have chosen $\mu_\phi=0.01$ and $\mu_l=0.46$ (requiring $b-\tau$ unification) and the initial values of the Dirac neutrino masses
\[m_{N_1}(M_R)=0.0031~{\rm GeV},~~m_{N_2}(M_R)=0.2825~{\rm GeV},~~m_{N_3}=71.86~{\rm GeV}\]
such that the masses evaluated at the GUT-scale, $m_{N_i}(M_G)$, agree with those predicted from the specific $SO(10)$ model
described in Section 6. For consistency check, we note that the $SO(10)$ model predicted eigenvalues of $M_D$ given by Eq.~(\ref{eq:mnuso10}),
\[m_{N_i}^{\rm predicted}=(0.0028,~0.2538,~77.8046)~{\rm GeV},\]
agree quite well with those obtained from the RGEs,
\[m_{N_i}^{\rm RG}(M_G) = (0.0028,~0.2538,~77.8106)~{\rm GeV}.\]

\end{document}